\documentclass{aa}  

\usepackage{graphicx}
\usepackage{amsmath}
\usepackage{xcolor}
\usepackage{txfonts}
\begin{document} 
\title{Comparing different mass estimators for a large subsample of the {\it Planck}-ESZ clusters}
\titlerunning{Comparing different mass estimators for the ESZ sample}

\author{
L. Lovisari\inst{1,2},
S. Ettori\inst{1,3},
M. Sereno\inst{1,3},
G. Schellenberger\inst{2},
W. R. Forman\inst{2},
F. Andrade-Santos\inst{2},     
\and
C. Jones\inst{2} 
}
\authorrunning{L. Lovisari et al.}

\institute{
INAF - Osservatorio di Astrofisica e Scienza dello Spazio di Bologna, via Piero Gobetti 93/3, I-40129 Bologna, Italia \\
\email{lorenzo.lovisari@inaf.it}
\and
Center for Astrophysics $|$ Harvard $\&$ Smithsonian, 60 Garden Street, Cambridge, MA 02138, USA
\and
INFN, Sezione di Bologna, viale Berti Pichat 6/2, I-40127 Bologna, Italia
}

\date{Received June 22, 2020; accepted October 6, 2000}

\abstract
{Total mass is arguably the most fundamental property for cosmological studies with galaxy clusters. The individual cluster masses can be obtained with different methods, each with its own biases and limitations. Systematic differences in mass measurements can strongly impact the determination of the hydrostatic bias and of the mass-observable relations, key requirements of many cluster abundance studies.}
{We investigate the present differences in the mass estimates obtained through independent X-ray, weak-lensing, and dynamical studies using a large subsample of the {\it Planck}-ESZ clusters. We also discuss the implications for mass bias analyses.}
{After assessing the systematic differences in the  X-ray-derived masses reported by distinct groups, we examine the mass estimates obtained with independent methods and quantify the differences as the mean ratio 1-$b$=M$_{\rm HE}$/M$_{\rm WL,dyn}$, where HE refers to hydrostatic masses obtained from X-ray observations, WL refers to the results of weak-lensing measurements, and dyn refers to the mass estimates either from velocity dispersion or from the caustic technique. So defined, the 1-$b$ parameter includes all possible astrophysical, observational, and methodological biases in one single value. }
{Recent X-ray masses reported by independent groups show average differences smaller than $\sim$10$\%$, posing a strong limit on the systematics that can be ascribed to the differences in the X-ray analysis when studying the hydrostatic bias. The mean ratio between our X-ray masses and the weak-lensing masses in the LC$^2$-single catalog is 1-$b$=0.74$\pm$0.06 corresponding to a mass bias of 26$\pm$6$\%$, a value insufficient to reconcile the {\it Planck} cluster abundance and CMB results. However, the mean mass ratios inferred from the WL masses of different projects vary by a large amount, with APEX-SZ showing a bias consistent with zero (1-$b$=1.02$\pm$0.12), LoCuSS  and CCCP/MENeaCS showing a significant difference (1-$b$=0.76$\pm$0.09 and 1-$b$=0.77$\pm$0.10, respectively), and WtG pointing to the largest deviation (1-$b$=0.61$\pm$0.12) which would substantially reduce the tension between the {\it Planck} results. Because of small differences between our M-Y$_{\rm X}$ relation and the one used by the {\it Planck} collaboration, our X-ray masses are on average 7$\%$ lower (4$\%$ at the same physical radius) than the {\it Planck} masses and can further reduce the required bias.
At odds with the WL results, the dynamical mass measurements show better agreement with the X-ray hydrostatic masses, although there are significant differences when relaxed or disturbed clusters are used. However, the comparison is currently limited by the small sample sizes. 
}
{The systematic differences between total masses obtained with recent independent X-ray analyses are smaller than those found in previous studies. This shifts the focus to WL and dynamical studies for a better convergence of the level of mass bias. However, the ratios obtained using different mass estimators suggest that there are still systematics that are not accounted for in all the techniques used to recover cluster masses. This prevents the determination of firm constraints on the level of hydrostatic mass bias in galaxy clusters.  
}


\maketitle

\section{Introduction} \label{sec:intro}
Galaxy cluster abundances provide sensitive constraints on the cosmological parameters that govern the expansion history of the Universe (e.g., \citealt{2009ApJ...692.1060V}, \citealt{2011ARA&A..49..409A}, \citealt{2012ARA&A..50..353K}). Typical galaxy cluster number count experiments infer total masses from a mass-observable scaling relation, which in turn requires calibration of the mass-proxy bias and scatter (e.g., see \citealt{2016A&A...594A..24P}, \citealt{2018ApJS..235...20H}, \citealt{2019ApJ...878...55B}, for SZ-based experiments). Thus, to use clusters for cosmological studies, accurate total masses for large samples of clusters with well-understood selection criteria are required.

One surprising result from the {\it Planck} collaboration is that $\sigma_8$ and $\Omega_M$  derived from Sunyaev-Zel'dovich (SZ, \citealt{1972CoASP...4..173S}) cluster counts are inconsistent with the values derived from the {\it Planck} cosmic microwave background (CMB) cosmology (see \citealt{2016A&A...594A..24P}). The tension would be removed if clusters were 60-85$\%$ more massive than estimates based on scaling relations. Although a redshift dependence could alleviate the tension (\citealt{2017MNRAS.468.3322S}), this level of bias is much larger than expected from numerical simulations which suggest that X-ray-derived masses, based on hydrostatic equilibrium, are biased low by up to 30$\%$ (e.g., \citealt{2008A&A...491...71P}, \citealt{2010A&A...514A..93M}, \citealt{2012NJPh...14e5018R}, \citealt{2014ApJ...791...96R}, \citealt{2016ApJ...827..112B}, \citealt{2020A&A...634A.113A}, \citealt{2020arXiv200111508B}).  
This level of bias is also not fully supported by observations, although different studies find quite different results, hampered by the different mass and redshift ranges and related selection effects explored in each analysis, which are also performed with different methodologies. For instance, assuming that weak-lensing (WL) re-construction provides unbiased measurements of the true mass, \cite{2014MNRAS.443.1973V} with a sample of N=22 clusters with mass and redshift ranges of M=[4-30]$\times$10$^{14}$M$_{\odot}$ and $z$=[0.15-0.7], found a bias (1-$b$=M/M$_{\rm true}$=0.69$\pm$0.07 which is consistent at the 1$\sigma$ level with the bias required by {\it Planck}. \cite{2017A&A...604A..89P} (N=21, M=[3-11]$\times$10$^{14}$M$_{\odot}$ and $z$=[0.19-0.89] found a similar result (i.e., 1-$b$=0.73$\pm$0.10). \cite{2020MNRAS.497.4684H} (N=100, M=[0.5-20]$\times$10$^{14}$M$_{\odot}$ and $z$=[0.05-0.55]) and \citet{2018PASJ...70S..28M} (N=5, M=[2-30]$\times$10$^{14}$M$_{\odot}$ and $z$=[0.06-0.33]) found a smaller bias (0.84$\pm$0.04 and 0.80$\pm$0.14, respectively), while the results of \cite{2019A&A...621A..39E} (N=13, M=[2-9]$\times$10$^{14}$M$_{\odot}$ and $z$=[0.05-0.09]), \cite{2016MNRAS.456L..74S} (N=50, M=[2-15]$\times$10$^{14}$M$_{\odot}$ and $z$=[0.15-0.3]), and \cite{2014A&A...564A.129I} (N=8, M=[1-5]$\times$10$^{14}$M$_{\odot}$ and $z$=[0.4-0.8])  point to a bias of only a few percent. Many other studies have investigated this subject and found a mass bias in the range above-mentioned (e.g., see \citealt{2010ApJ...711.1033Z}, \citealt{2013ApJ...767..116M}, \citealt{2014MNRAS.442.1507G}, \citealt{2015MNRAS.449..685H}, \citealt{2016MNRAS.457.1522A}, \citealt{2019ApJ...875...63M}). Thus, while assessing the level of mass bias in galaxy cluster is a fundamental step for their application to precision cosmology, the magnitude of the bias is still an open issue.

Under the assumption of hydrostatic equilibrium, the masses for individual clusters of galaxies can be derived from the distribution of their X-ray emitting gas. However, for most of the clusters the hydrostatic equilibrium assumption is probably invalid, at least mildly, making hydrostatic mass determinations more prone to biases. The more disturbed the cluster is, the more underestimated the X-ray cluster mass is expected to be.  In fact, the major source of bias is associated with bulk gas motions which act as additional pressure support against gravity (e.g., \citealt{2006MNRAS.369.2013R}, \citealt{2008A&A...491...71P}, \citealt{2009ApJ...705.1129L}, \citealt{2012NJPh...14e5018R}, \citealt{2016ApJ...827..112B}, \citealt{2018MNRAS.481L.120V}). On top of that, the presence of temperature inhomogeneites and/or multi-phase gas can bias the estimate from X-ray instruments because of the higher weight of the colder gas components (e.g., see \citealt{mazzotta+04}). Also, the bias is found to be higher in the outskirts, because of the increasing contribution of gas motions and temperature inhomogeneities (e.g., \citealt{2012NJPh...14e5018R}).

At the moment, WL is considered the most reliable method to determine accurate masses because it measures the total mass directly, without relying on baryonic tracers. Hence, the lensing signal does not require any assumption about the dynamical state of the gravitating mass of the cluster.  Motivated by the results of numerical simulations (e.g., \citealt{2011ApJ...740...25B}, \citealt{2011MNRAS.414.1851O}, \citealt{2012NJPh...14e5018R}) showing that the average weak-lensing masses are biased low by less of 10$\%$, many cluster mass studies used WL measurements to constrain cluster masses. Nonetheless, individual systems can suffer significant biases depending on the observer's viewing angle with respect to the cluster mass distribution and the presence of substructures (see, e.g., \citealt{2010A&A...514A..93M}, \citealt{2018ApJ...860L...4S}, and \citealt{2019SSRv..215...25P} for more details). Moreover, the triaxial distribution of mass introduces a scatter of up to 30$\%$ in lensing masses for individual clusters. This quite large scatter implies that samples of a few dozens of clusters are required to constrain the normalization of scaling relations to better than 10$\%$ (e.g., \citealt{2011ApJ...740...25B}).

Techniques based on galaxy dynamics, such as the caustic technique (e.g., \citealt{1997ApJ...481..633D},  \citealt{1999MNRAS.309..610D}) and methods using either the Jeans equation or the virial theorem (e.g., \citealt{2013A&A...558A...1B}, \citealt{2017ApJ...844..101A}), can also be used to estimate total masses, although these methods are impeded by the expensive observational requirements to obtain spectroscopic measurements of galaxies velocities.
Similar to WL estimates, the total masses obtained with the caustic technique do not rely on the hypothesis of dynamical equilibrium which is instead required by the Jeans and virial estimators.
In the literature, there are fewer caustic studies than those using WL measurements to compare hydrostatic and dynamical masses for relatively large samples.  \cite{2016MNRAS.461.4182M} found a small or zero value  of the hydrostatic bias if the caustic masses are assumed to be equivalent to the true mass, and \cite{2016ApJ...819...63R} found a similar result comparing SZ mass estimates (calibrated with hydrostatic X-ray masses) with the caustic masses. On the contrary, \cite{2017A&A...606A.122F} found that the dynamical masses are on average larger than the hydrostatic values with the ratio between the Jeans, virial theorem, and caustic estimators, with respect to the hydrostatic mass, being 1-$b$=1.22$\pm$0.18, 1-$b$=1.51$\pm$0.26, and 1-$b$=1.32$\pm$0.18, respectively.

Each of the above-mentioned approaches requires various hypotheses and suffers from different systematics (see, e.g., the review by \citealt{2019SSRv..215...25P} and references therein). Thus,  an inter-comparison of the different results can provide useful information to better understand their respective systematics. In this paper, we examine the differences in such measurements for a large sample of {\it Planck}-selected clusters analyzed in X-rays in \citet[hereafter L17]{2017ApJ...846...51L} and \citet[hereafter L20]{2020ApJ...892..102L}.  The comparison will also shed light on the magnitude of the hydrostatic equilibrium mass bias, and whether the value 1-$b$=0.58$\pm$0.04, required to reconcile the {\it Planck} CMB with number counts (\citealt{2016A&A...594A..24P}, see also \citealt{2019A&A...626A..27S}), is consistent with a significant underestimate of the X-ray-derived masses. In fact, the Planck team estimated cluster masses using the SZ signal Y$_{\rm SZ}$\footnote{The Y$_{\rm SZ}$ parameter corresponds to the Compton y-profile integrated within a sphere of radius R$_{500}$, the radius within which the mean overdensity of a cluster is 500 times the critical density of the Universe at the cluster redshift.} calibrated against hydrostatic masses. Therefore, a significant bias would help to reconcile the Planck results.

The paper is organized as follows. In Section 2, we introduce our sample and describe the data analysis. In Section 3, we present our results, and, in Section 4, we discuss our findings. A summary is provided in Section 5.
In Appendix \ref{estimators} we compare the biases computed with different estimators, and in Appendix \ref{Mxray} we show the comparison between the X-ray masses derived in different studies.

Throughout this paper, we assume a flat $\Lambda$CDM cosmology with $\Omega_m$=0.3 and H$_0$=70 km/s/Mpc. Uncertainties are at the 68$\%$ confidence level. Log is always base 10.

\section{Data and analysis}

\subsection{Sample}
The {\it Planck} Early Sunyaev-Zel'dovich (ESZ, \citealt{2011A&A...536A...8P}) sample consists of 188 massive clusters, which were selected by imposing a signal-to-noise ratio (SNR) threshold of 6 and a Galactic cut $|{\rm b}|$$>$14$^\circ$ on the catalog of SZ detections from the first ten months of the {\it Planck} survey. Although SZ-selected samples suffer from Malmquist bias and are not fully mass-selected, they are considered  representative of the underlying cluster population. In fact, simulations have shown that SZ quantities do not strongly depend on the dynamical state of the clusters (e.g., \citealt{2005ApJ...623L..63M}), with only a modest effect due to mergers (e.g., \citealt{2012ApJ...758...74B}). This is supported observationally by the tight correlation between SZ signal and mass (e.g., \citealt{2011A&A...536A..11P}). Moreover, the morphology of the source has little impact on the detection procedure in the Planck survey, as shown via MonteCarlo simulations (\citealt{2016A&A...594A..27P}).

L17 and L20 analyzed the {\it XMM-Newton} data to determine the morphological and global cluster properties  for a representative subsample  of 120 ESZ clusters,  selected to ensure that R$_{500}$ is completely covered by {\it XMM-Newton} observations (either single or multiple pointings). A single pipeline was used in the analysis allowing a uniform processing of the data, eliminating  one source of uncertainties in the comparison with the masses estimated from other methods.

For all these 120 clusters, we know the dynamical state and both X-ray and SZ masses (from L20 and \citealt{2016A&A...594A..27P}, respectively). 
Instead of the SZ masses estimated in \cite{2011A&A...536A...8P}, we preferred to use the more recent and more accurate PSZ2 masses reported in \cite{2016A&A...594A..27P} obtained from the 29 month full {\it Planck} mission data. We note that 5 ESZ clusters are not included in the PSZ2 catalog because they fall into the PSZ2 point source mask, but only 3 of these overlap with our subsample of 120 clusters. For these clusters we use the PSZ1 masses. 

\begin{figure}[t!]
\centering
\includegraphics[width=0.47\textwidth]{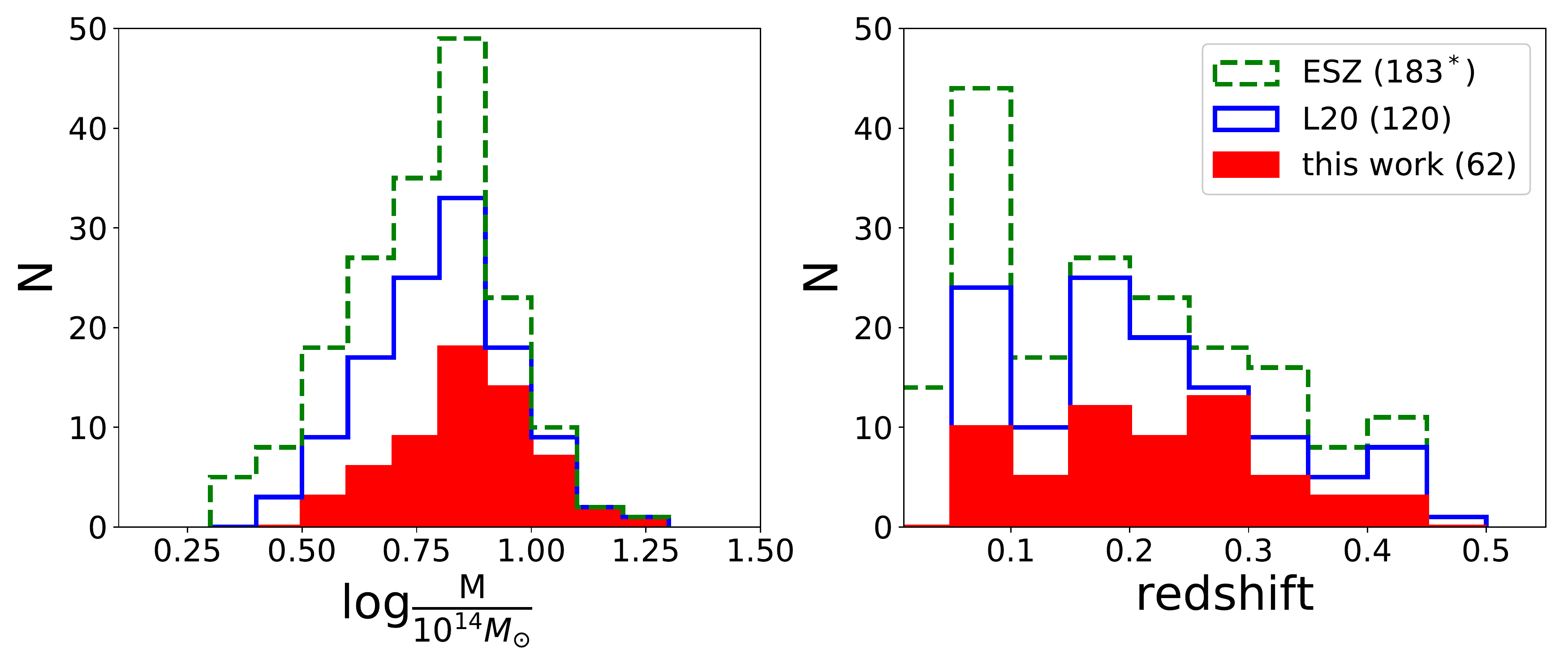}
\caption{Distribution of the cluster masses ({\it left panel}) and redshifts ({\it right panel}) for the ESZ clusters (dashed green), the L20 sample (solid blue), and the clusters with weak-leansing masses (filled red). Numbers in the legend represent the sample size.}
\label{fig:samples}
\end{figure}

For WL masses, we used the compilation obtained in the CoMaLit framework (see, e.g., \citealt{2015MNRAS.450.3633S}). The Literature Catalogs of Weak Lensing Clusters (LC$^2$, see \citealt{2015MNRAS.450.3665S}) is publicly available at \url{http://pico.oabo.inaf.it/~sereno/CoMaLit/LC2/}. 
LC$^2$-single (v. 3.9) is a meta-catalog with 806 unique entries. When multiple analyses per cluster are available, preference is given to studies exploiting the deeper observations and multi-band optical coverage for optimal galaxy background selection. There are 62 clusters in common with the sample in L20. The distribution of mass and redshift for these clusters with respect to the ESZ and L20 samples is shown in Fig. \ref{fig:samples}. 

Given the relatively large sample, we also investigated whether or not the X-ray/SZ masses of disturbed clusters are significantly more biased than the masses derived for relaxed clusters. The dynamical state of each cluster was determined by using two morphological parameters: one sensitive to the core-properties (i.e., surface brightness concentration) and one sensitive to the large scale inhomogeneities (i.e., centroid-shift). In  L20 (see also \citealt{2013AstRv...8a..40R}) these two parameters have been combined to define a new general parameter M (hereafter M$_{\rm par}$), which allows us to use a single value to rank the clusters from the most relaxed to the most disturbed in the sample of interest. In the following, we refer to relaxed and disturbed cluster when M$_{\rm par}$ is greater than and smaller than zero, respectively. This classification provides a simple way to divide a sample between the most relaxed and most disturbed clusters but does not necessarily provide an absolute reference.

\subsection{Estimation of the hydrostatic bias}
We estimated the mass ratio by fitting the data with LIRA which allows normalization, slope, and scatters (and relative uncertainties) to be fitted simultaneously\footnote{\label{lira}LIRA (LInear Regression in Astronomy) is based on a Bayesian method (see \citealt{2016MNRAS.455.2149S} for more details) and is available from the R archive network at \url{https://cran.r-project.org/web/packages/lira/index.html}.}.

In particular, for each subsample of WL (or dynamical) masses we linearly fit our data as
\begin{gather}
\log\left(\frac{\rm M_{HE}}{\rm 6\times10^{14}M_{\odot}}\right)=\alpha+\beta\log\left(\frac{\rm M_{true}}{\rm 6\times10^{14}M_{\odot}}\right)\pm\sigma_{\rm HE|true}\\
\nonumber{\log{(\rm M_{true})}=\log{\rm ( M_{WL})}\pm\sigma_{\rm WL|true}}
\end{gather}
where $\alpha$ quantifies the mass ratio (i.e., 1-$b$=M$_{\rm HE}$/M$_{\rm WL}$=10$^{\alpha}$), $\beta$ is fixed to 1 (unless otherwise specified), and the M$_{\rm HE}$ and M$_{\rm WL}$ masses are evaluated at their respective R$_{500}$ (unless otherwise stated). M$_{\rm true}$ represents the unbiased cluster mass. The intrinsic scatters $\sigma_{\rm HE|true}$ and $\sigma_{\rm WL|true}$ are assumed to be uncorrelated.
As described in \cite{2016MNRAS.455.2149S}, for each set of parameters, the linear regression is performed calling the function {\it lira}, whose output are Markov chains produced with a Gibbs sampler. The median values of $\alpha$, $\sigma_{\rm HE|true}$, $\sigma_{\rm WL|true}$ and their 1$\sigma$ uncertainties are taken from the distribution of 80,000 Markov chains. 

For comparison, in Appendix \ref{estimators} we provide also the bias estimated by computing the geometric mean and the median.

\subsection{X-ray total masses}\label{sectMxray}
Under the assumption of spherical symmetry, one can solve the hydrostatic equilibrium equation and recover the total mass with only two ingredients: gas density and temperature profiles. For the ESZ sample, the gas density profiles were recovered  from the best-fit parameters of a double $\beta$-model, while, for the temperature profiles, we used the 3D model by \cite{2006ApJ...640..691V} projected along the line of sight to fit the observed data points following the implementation presented in \cite{2017MNRAS.469.3738S}.

\begin{figure}[t!]
\centering
\includegraphics[width=0.45\textwidth]{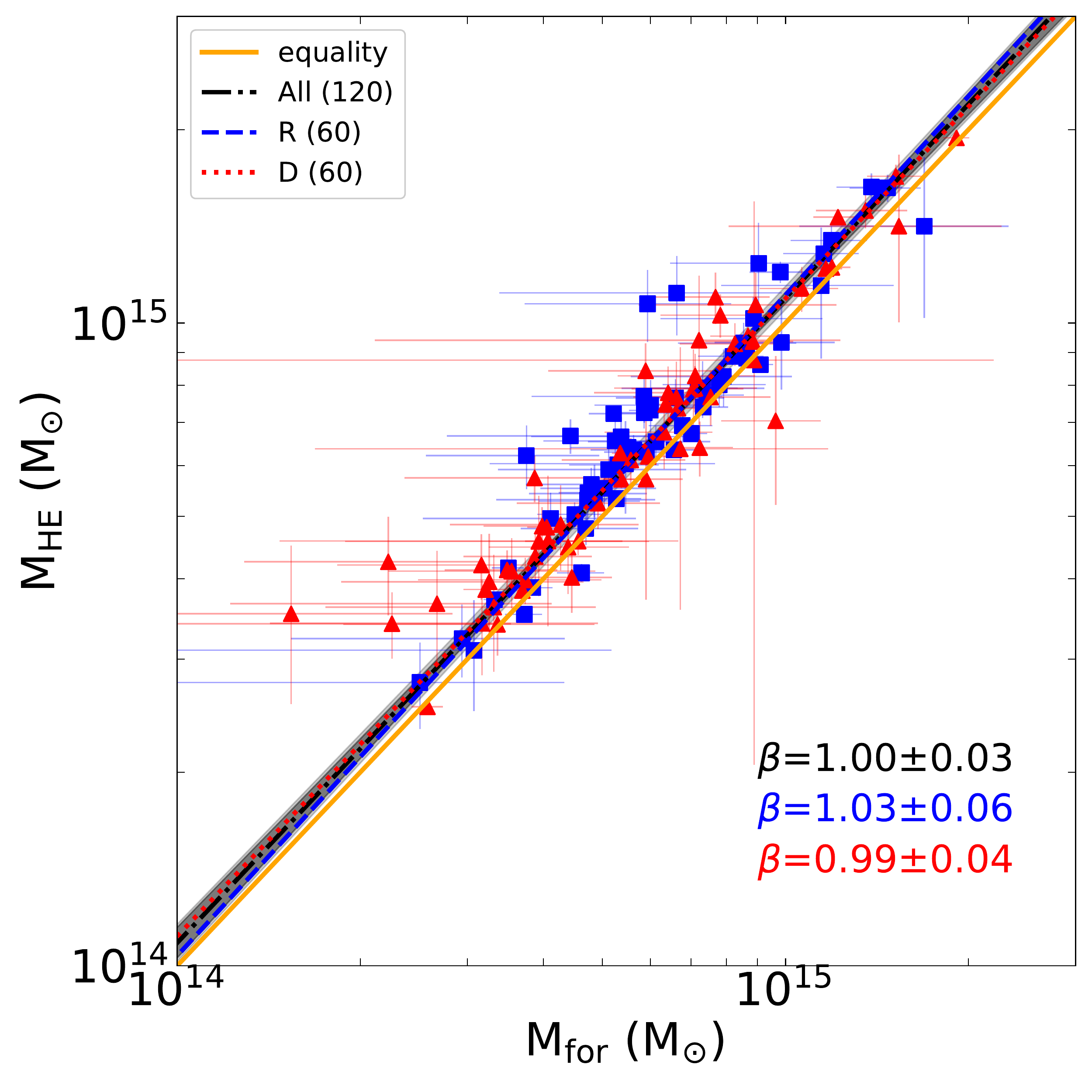}
\caption{Comparison between the total masses obtained with a forward approach (M$_{\rm for}$) and  a backward approach (M$_{\rm HE}$). Relaxed (R) and disturbed (D) clusters are displayed as blue squares and red triangles, respectively. Numbers in the legend represent the sample size. The $\beta$ values represent the fitted slope for disturbed (red), relaxed (blue), and all (black) clusters, respectively.}
\label{fig:HEhe}
\end{figure}

In L20, the X-ray total cluster masses have been obtained by assuming a Navarro-Frenk-White (NFW, \citealt{1996ApJ...462..563N,1997ApJ...490..493N}) model for the mass profile with the relation from \cite{2013ApJ...766...32B} between the Dark Matter concentration and radius as a prior. However, we also estimated the masses by directly solving the equation of hydrostatic equilibrium without any prior on the form of the gravitational potential. Following the nomenclature by \cite{2013SSRv..177..119E} (see also \citealt{2019SSRv..215...25P}), we refer to these masses as backward (hereafter M$_{\rm HE}$) and forward (hereafter M$_{\rm for}$), respectively, each measured at their R$_{500}$.

There is a good correlation between M$_{\rm HE}$  and M$_{\rm for}$ (see Fig.  \ref{fig:HEhe}), with a one-to-one relation and small intrinsic scatter with respect to the fitted relation (i.e., 4$\pm$1$\%$). However, the M$_{\rm for}$ masses are on average $\sim$9$\%$ lower than M$_{\rm HE}$ masses, almost independently of the dynamical state of the clusters. Instead, the mass difference slightly decreases as a function of the X-ray data quality. For instance, clusters for which the last temperature bin is at a radius $<$0.6R$_{500}$, within 0.6-0.8R$_{500}$, or $>$0.8R$_{500}$ have an average offset of 13$\%$, 9$\%$, and 7$\%$, respectively. Thus, the smaller the extrapolation required, the better the two estimates agree. Low redshift clusters (i.e., z$<$0.2) show a marginally smaller offset (7$\%$ on average) than high redshift (i.e., z$>$0.2) clusters (11$\%$). This is mainly due to higher quality data of the low redshift clusters that allow measurements of the temperature to a larger fraction of R$_{500}$. 

Previous studies showed that X-ray masses obtained with different analyses can be quite discordant. Indeed, this is a source of noise in the calculation of the level of hydrostatic bias. Since most of our clusters have been previously studied by other authors, here we provide a comparison of the mass estimates, evaluated at their respective R$_{500}$.

There are 48 clusters that overlap with the sample used by the \cite{2011A&A...536A..11P} to investigate the relation between the integrated Compton parameter and several X-ray-derived properties. On average, our masses are 4$\pm$2$\%$ lower than their masses estimated using the M-Y$_{\rm X}$ relation\footnote{The Y$_{\rm X}$=M$_{\rm gas}\times$kT parameter measures the total gas energy content and is the X-ray counterpart of the Sunyaev-Zel'dovich signal Y$_{\rm SZ}$.} by \cite{2010A&A...517A..92A}. There is a mild, not statistically significant, dependence on the dynamical state: for the relaxed clusters, our mass estimates are in agreement  (i.e., offset of 2$\pm$2$\%$) with \cite{2011A&A...536A..11P},  while for disturbed clusters the average offset is slightly larger (i.e., offset of 7$\pm$3$\%$).

There are 18 clusters in common with \cite{2014MNRAS.443.2342M}, who analyzed 50 clusters from the LoCuSS \citep[Local Cluster Substructure Survey, e.g.][]{2008A&A...482..451Z} sample. Of these 18 clusters, \cite{2014MNRAS.443.2342M} provide 17 (14) total masses obtained with {\it Chandra} ({\it XMM-Newton}) data. The mass profiles in \cite{2014MNRAS.443.2342M} were derived with a forward approach using a powerlaw model for the temperature profile and the parametric form by \cite{2006ApJ...640..691V} for the gas density profile. On average, our masses are 7$\pm10\%$ (10$\pm8\%$) lower than what they reported. 

There are 12 clusters in common with \cite{2017MNRAS.469.3738S} who analyzed the HIFLUGCS  \citep[HIghest X-ray FLUx Galaxy Cluster Sample,][]{2002ApJ...567..716R} sample with {\it Chandra} data. The mass profiles were derived with a forward approach using a double $\beta$-model for the density profile and several parametric forms which approximate a powerlaw in the outer regions, for the temperature profiles. Our total masses are, on average, 9$\pm11\%$ lower.

There are 6 clusters in common with the CLASH \citep[Cluster Lensing And Supernova survey with Hubble,][]{2012ApJS..199...25P} sample. The masses were obtained with a backward approach assuming an NFW form for the mass profile and a triple $\beta$-model for the density profiles.
Our total masses are on average $\sim$5$\%$ lower than the masses estimated by  \cite{2014ApJ...794..136D} with {\it Chandra} data and the difference reduces to $\sim$2$\%$ when compared to the masses obtained with {\it XMM-Newton} data.   

Finally, there are 5 clusters in common with XCOP (XMM Cluster Outskirts Project, \citealt{2017AN....338..293E}). The total masses derived in L20 are on average 11$\%$ lower than the masses estimated with the backward method and an NFW model as reported by \citet{2019A&A...621A..39E}. For 2 clusters (i.e., A3266 and ZwCl1215.1+0400) our total masses are $\sim$30$\%$ lower than the XCOP estimate, but are in agreement with the {\it Planck} masses. Of the XCOP clusters, these two are the ones that deviate more significantly with respect to the {\it Planck} estimates.  

In Appendix \ref{Mxray} we show the plot comparing the L20 total masses with those from literature.

\begin{table}[t!]
\centering
\caption{Summary of the 1-$b$=M$_{\rm HE}$/M$_{\rm SZ}$, estimated at R$_{500}$. For each subsample we provide the number of clusters, the median SZ masses (i.e., ${\rm \tilde{M}}_{SZ}$), and the median redshift (i.e., $\tilde{z}$). The intrinsic scatters are given in percent.}
\label{tab:xsz}
\resizebox{0.49\textwidth}{!}{
\begin{tabular}{c c c c c c c}
\hline
\hline
\noalign{\smallskip}
\multicolumn{7}{c}{All} \\
\noalign{\smallskip}
\hline
\noalign{\smallskip}
Sample & N & ${\rm \tilde{M}}_{SZ}$ & $\tilde{\rm z}$ & 1-$b$ & $\sigma_{SZ}$ & $\sigma_{HE}$ \\
\noalign{\smallskip}
\hline
\noalign{\smallskip}
L20 All & 117 & 6.59 & 0.19 & 0.93$\pm$0.02 & 5$\pm$3 & 16$\pm$2  \\
\noalign{\smallskip}
L20 R & 57 & 6.68 & 0.20 & 0.96$\pm$0.03 & 7$\pm$4 & 16$\pm$3  \\
\noalign{\smallskip}
L20 D & 60 & 6.51 & 0.19 & 0.90$\pm$0.03 & 6$\pm$4 & 15$\pm$3 \\
\noalign{\smallskip}
LC$^2$-{\it single} & 60 & 7.56 & 0.22 & 0.93$\pm$0.02 & 5$\pm$3 & 17$\pm$3 \\
\noalign{\smallskip}
CCCP/MENeaCS & 25 & 7.65 & 0.18 & 0.94$\pm$0.04 & 11$\pm$6 & 14$\pm$6  \\
\noalign{\smallskip}
APEX-SZ & 19 & 8.76 & 0.28 & 0.95$\pm$0.04 & 4$\pm$3 & 15$\pm$3  \\
\noalign{\smallskip}
LoCuSS & 18 & 7.78 & 0.22 & 0.88$\pm$0.04 & 6$\pm$4 & 13$\pm$4  \\
\noalign{\smallskip}
WtG  & 17 & 8.27 & 0.29 & 0.92$\pm$0.04 & 5$\pm$3 & 13$\pm$4  \\
\noalign{\smallskip}
\hline
\end{tabular}
}
\end{table}

\begin{figure}[h!]
\centering
\includegraphics[width=0.45\textwidth]{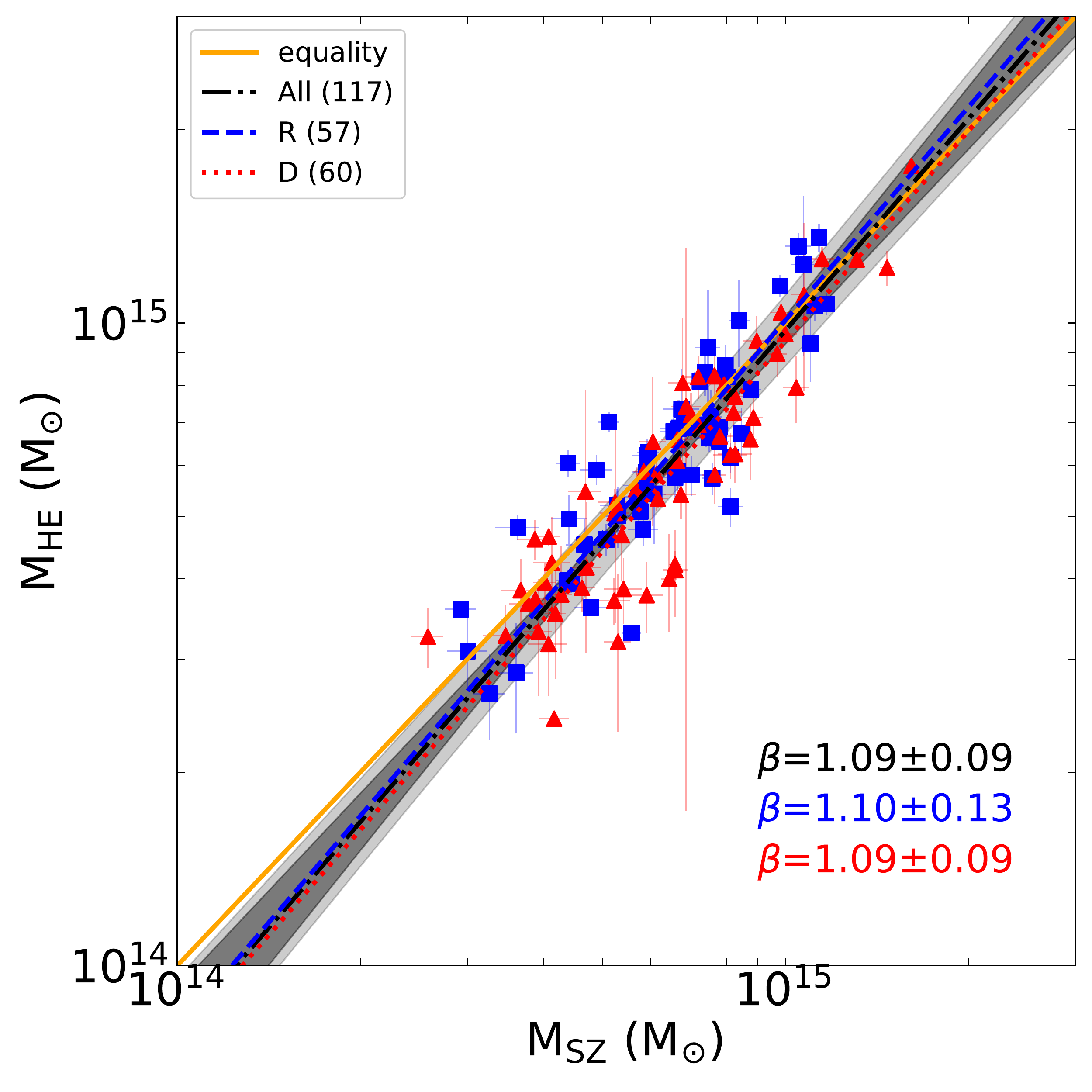}
\caption{M$_{\rm HE}$-M$_{\rm SZ}$ relation for the ESZ clusters investigated in L20. Relaxed (i.e., M$_{\rm par}$$>$0) and disturbed (i.e., M$_{\rm par}$$<$0) clusters are shown in blue squares and red triangles, respectively. The dark shaded area represent the 1$\sigma$ statistical error for the full sample.}
\label{fig:HEsz}
\end{figure}

\subsection{{\rm M$_{\rm HE}$-M$_{\rm SZ}$}}\label{sect:MheMsz}
The thermal SZ effect directly measures the line-of-sight pressure, and thus, in the adiabatic scenario, the thermal energy content of the cluster gas, which is strongly correlated with the total mass. However, the SZ-derived masses largely rely on prior information on the calibration of the integrated SZ Comptonization (i.e., Y$_{\rm SZ}$) with cluster masses from X-ray and/or weak-lensing analyses. The {\it Planck} masses have been determined using the X-ray hydrostatic masses to calibrate the Y$_{\rm X}$-Y$_{\rm SZ}$ relation. Therefore, the comparison M$_{\rm HE}$ and M$_{\rm SZ}$ indirectly provides the comparison between our ESZ masses (i.e., M$_{\rm HE}$) and the hydrostatic masses used by Planck. 

In Fig. \ref{fig:HEsz} we show the comparison between the X-ray (i.e., M$_{\rm HE}$) and PSZ2 masses (i.e., M$_{\rm SZ}$). The SZ masses are, on average, 7$\%$ higher than the X-ray masses (see Table \ref{tab:xsz}). When we compute the X-ray masses at R$_{500}$ determined by {\it Planck}, the difference decreases to $\sim$4$\%$. When left free to vary, the slope of the M$_{\rm HE}$-M$_{\rm SZ}$ relation is found to be only slightly steeper than the equality line (i.e., $\beta$=1.09$\pm$0.09) indicating a not significant mass dependence. There is no variation in slope when the sample is split based on the dynamical state of the clusters (see blue and red lines in Fig. \ref{fig:HEsz}). Even when selecting the most relaxed (top 30$\%$) and most disturbed clusters in our sample, we do not find any significant dependence. However, there is a mild effect on the normalization with relaxed clusters more in agreement with the {\it Planck} masses than the disturbed clusters.

When splitting the sample in 3 redshift bins (i.e., $z$$<$0.15, 0.15$<$$z$$<$0.25, $z$$>$0.25), we found the M$_{\rm HE}$/M$_{\rm SZ}$ ratio to be 0.96$\pm$0.04, 0.90$\pm$0.03, and 0.96$\pm$0.04, respectively. Thus, there are no indication of a dependence on redshift, as also confirmed by the Spearman rank test (i.e., r=0.06 and p=0.53). We also subdivided each redshift bin based on cluster mass but we did not find any trend. 

We also investigated if any of the WL subsamples show a significantly different M$_{\rm HE}$/M$_{\rm SZ}$ ratio. The LoCuSS subsample shows the lowest value (i.e., 1-$b$=0.88$\pm$0.04) but it is still consistent within 1$\sigma$ with the results from the other subsamples.

\section{Results}
Cluster masses can be directly estimated through X-ray observations, gravitational weak-lensing, and dynamical analysis of cluster galaxies, or  alternatively using a mass proxy, such as the SZ flux which relies on prior information on the  SZ-mass calibration obtained from X-ray or weak-lensing masses. Although each method requires some assumptions and is limited by different systematics, weak-lensing and caustic technique do not require the assumption of hydrostatic equilibrium, and numerical studies suggest that they can provide cluster mass estimates with little bias. In the following we compare our ESZ masses (i.e., M$_{\rm HE}$) with the ones from other mass estimators.

\subsection{X-ray vs weak-lensing masses} \label{sect:xwl}
\begin{figure}[t!]
\centering
\includegraphics[width=0.45\textwidth]{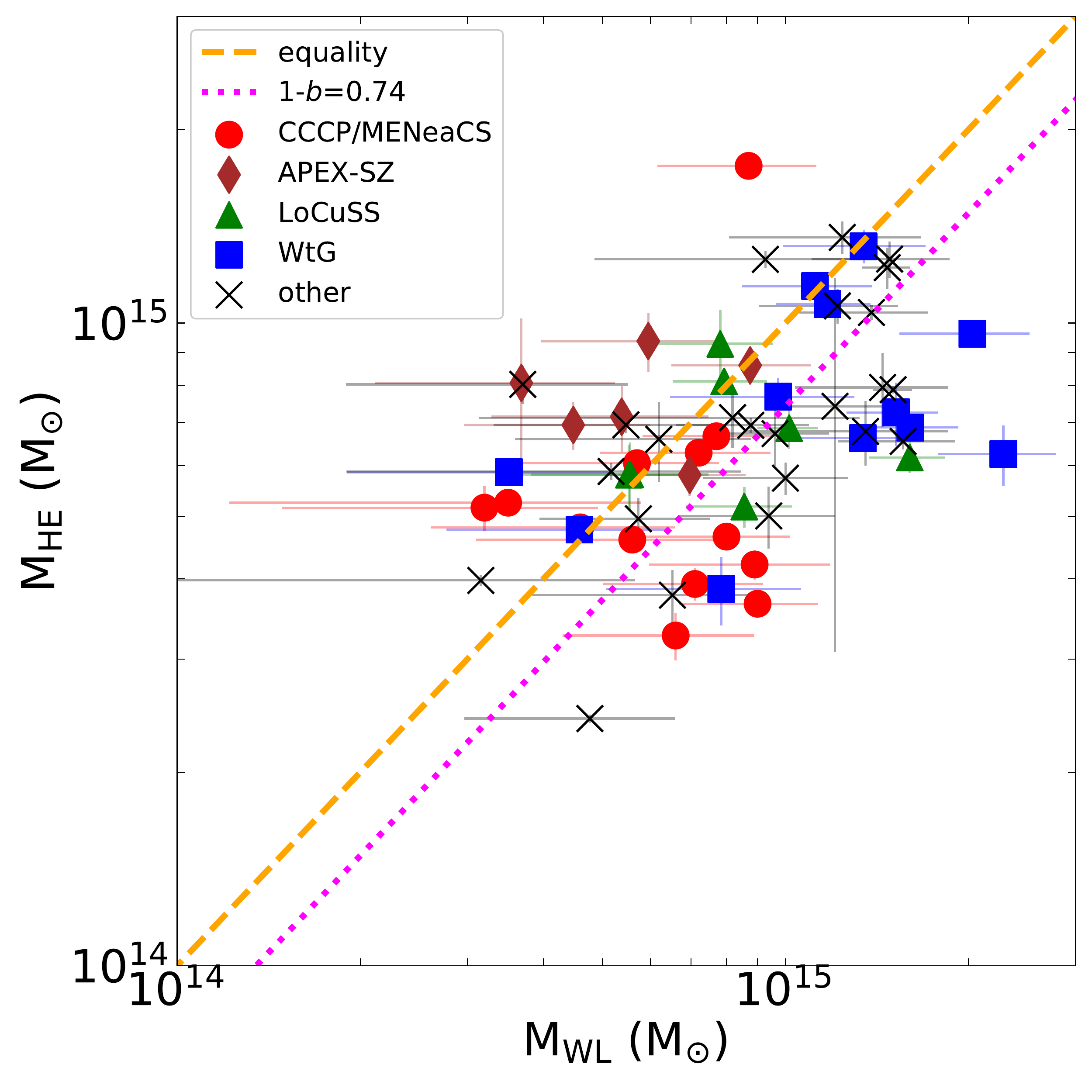}
\caption{Comparison between X-ray and WL masses from LC$^2$-single. Red circles, brown diamonds, green triangles, and blue squares represent the masses obtained by \cite{2020MNRAS.497.4684H}, \cite{2019MNRAS.488.1704K}, \cite{2016MNRAS.461.3794O},  and \cite{2014MNRAS.439...48A}, respectively. The magenta line represents the average M$_{\rm HE}$/M$_{\rm WL}$ ratio.}
\label{fig:HElen}
\end{figure}

In this section, we compare X-ray and WL masses estimated within R$_{500}$ which is one of the most common overdensities used for cosmological studies. In Fig. \ref{fig:HElen} we show the M$_{\rm HE}$-M$_{\rm WL}$ relation, while in Table \ref{tab:xszwl}, we provide the values 1-$b$ computed for all the investigated subsamples.

When the LC$^2$-single catalog is used, we find that the average ratio between HE and WL masses is 1-$b$=0.74$\pm$0.06. This number is almost independent of the dynamical state: relaxed and disturbed clusters have a mass ratio of 1-$b$=0.75$\pm$0.08 and 1-$b$=0.73$\pm$0.12, respectively.  By splitting the sample in three redshift bins, we find no dependence of the mass ratio on redshift (i.e., 1-$b$=0.72$\pm$0.09 for $z$$<$0.17, 1-$b$=0.75$\pm$0.12 for 0.17$<$$z$$<$0.29, and 1-$b$=0.76$\pm$0.12 for $z$$>$0.29). Instead, we find indication for a dependence on the total mass with smaller 1-$b$ values for high mass clusters and a better agreement for low mass systems. In detail, for M$_{\rm WL}$$\le$6.5$\times$10$^{14}$M$_{\odot}$ we found 1-$b$=1.15$\pm$0.11, for 
6.5$\times$10$^{14}$M$_{\odot}$$<$M$_{\rm WL}$$\le$10$^{15}$M$_{\odot}$ we found 1-$b$=0.73$\pm$0.11, and for M$_{\rm WL}$$>$10$^{15}$M$_{\odot}$ we found  1-$b$=0.62$\pm$0.08. This is also confirmed by the Spearman test which gives a significant anti-correlation (i.e., r=-0.67, p$<$0.01). 

Also for WL there are significant differences between masses reported by distinct groups (e.g., see also \citealt{2014ApJ...795..163U}, \citealt{2015MNRAS.450.3633S}). 

There are 27 clusters in common with the sample analysed by \citet[hereafter H20]{2020MNRAS.497.4684H} who determined the masses for 48 clusters in the Multi Epoch Nearby Cluster Survey (MENeaCS), and 52 clusters from the Canadian Cluster Comparison Project (CCCP). We find M$_{\rm HE}$/M$_{\rm WL}$=0.77$\pm$0.10. Among the common clusters, there are 17 relaxed clusters (i.e., M$_{\rm par}$$>$0) for which we obtain M$_{\rm HE}$/M$_{\rm WL}$=0.83$\pm$0.12 and 10 disturbed clusters (i.e., M$_{\rm par}$$<$0) for which we obtain M$_{\rm HE}$/M$_{\rm WL}$=0.73$\pm$0.21. Given the large uncertainties, the tendency of a better agreement for relaxed clusters is only marginal. The correlation between bias and dynamical state measured by M$_{\rm par}$ is also not supported by the Spearman test (i.e., r=-0.05, p=0.81).

There are 19 clusters in common with \citet[hereafter K19]{2019MNRAS.488.1704K} who presented the WL analysis for clusters from the APEX-SZ survey (see also \citealt{2019MNRAS.488.1728N}). Our HE masses are in agreement (i.e., M$_{\rm HE}$/M$_{\rm WL}$=1.02$\pm$0.12) with the WL masses by K19. 

There are 18 clusters in common with \citet[hereafter O16]{2016MNRAS.461.3794O}, who analyzed the LoCuSS clusters. The average mass ratio is 1-$b$=0.76$\pm$0.09. 
It is worth mentioning that LoCuSS shows the larger deviation in M$_{\rm HE}$/M$_{\rm SZ}$ (i.e., 12$\%\pm$4 offset, see Sect \ref{sect:MheMsz}). 

\begin{table}[t!]
\centering
\caption{Summary of the 1-$b$=M$_{\rm HE}$/M$_{\rm WL}$ values, estimated at R$_{500}$. The data for the CCCP/MENeaCS, APEX-SZ, LoCuSS, and WtG samples have been taken from H20, K19, O16, and A14, respectively. For each subsample we provide the number of clusters, the median WL mass (i.e., ${\rm \tilde{M}}_{WL}$), and the median redshift (i.e., $\tilde{z}$). Masses are in units of 10$^{14}$M$_{\odot}$. The intrinsic scatters are given in percent.}
\label{tab:xszwl}
\resizebox{0.49\textwidth}{!}{
\begin{tabular}{c c c c c c c}
\hline
\hline
\noalign{\smallskip}
\multicolumn{7}{c}{All} \\
\noalign{\smallskip}
\hline
\noalign{\smallskip}
Sample & N & $\tilde{\rm M}_{WL}$ & $\tilde{\rm z}$ & 1-$b$ & $\sigma_{WL}$ & $\sigma_{HE}$ \\
\noalign{\smallskip}
\hline
\noalign{\smallskip}
LC$^2$-{\it single} & 62 & 8.37 & 0.21 & 0.74$\pm$0.06 & 27$\pm$8 & 30$\pm$6 \\
\noalign{\smallskip}
CCCP/MENeaCS & 27 & 7.40 & 0.18 & 0.77$\pm$0.10 & 10$\pm$9 & 40$\pm$8  \\
\noalign{\smallskip}
APEX-SZ & 19 & 8.51 & 0.28 & 1.02$\pm$0.12 & 37$\pm$10 & 28$\pm$10 \\
\noalign{\smallskip}
LoCuSS & 18 & 8.24 & 0.22 & 0.76$\pm$0.09 & 23$\pm$9 & 24$\pm$8 \\
\noalign{\smallskip}
WtG  & 17 & 11.73 & 0.29 & 0.61$\pm$0.12 & 31$\pm$15 & 26$\pm$11  \\
\multicolumn{7}{c}{}\\
\hline
\hline
\noalign{\smallskip}
\multicolumn{7}{c}{Relaxed} \\
\noalign{\smallskip}
\hline
\noalign{\smallskip}
Sample & N & $\tilde{\rm M}_{WL}$ & $\tilde{\rm z}$  & 1-$b$ & $\sigma_{WL}$ & $\sigma_{HE}$ \\
\noalign{\smallskip}
\hline
\noalign{\smallskip}
LC$^2$-{\it single} & 34 & 8.65 & 0.21 & 0.75$\pm$0.08 & 27$\pm$9 & 18$\pm$8 \\
\noalign{\smallskip}
CCCP/MENeaCS & 17 & 7.10 & 0.17 & 0.83$\pm$0.12 & 21$\pm$13 & 25$\pm$10 \\
\noalign{\smallskip}
APEX-SZ & 11 & 8.75 & 0.27 & 0.90$\pm$0.17 & 42$\pm$14 & 27$\pm$13 \\
\noalign{\smallskip}
LoCuSS & 14 & 9.42 & 0.22 & 0.74$\pm$0.10 & 20$\pm$10 & 26$\pm$8  \\
\noalign{\smallskip}
WtG  & 10 & 11.46 & 0.23 & 0.71$\pm$0.17  & 22$\pm$18  & 32$\pm$14  \\
\multicolumn{7}{c}{}\\
\hline
\hline
\noalign{\smallskip}
\multicolumn{7}{c}{Disturbed} \\
\noalign{\smallskip}
\hline
\noalign{\smallskip}
Sample & N & $\tilde{\rm M}_{WL}$ & $\tilde{\rm z}$ & 1-$b$ & $\sigma_{WL}$ & $\sigma_{HE}$ \\
\noalign{\smallskip}
\hline
\noalign{\smallskip}
LC$^2$-{\it single} & 28 & 8.09 & 0.29 & 0.73$\pm$0.12 & 31$\pm$14 & 45$\pm$10 \\
\noalign{\smallskip}
CCCP/MENeaCS & 10 & 8.35 & 0.19 & 0.73$\pm$0.21 & 10$\pm$13  & 61$\pm$17 \\
\noalign{\smallskip}
APEX-SZ & 8 & 7.57 & 0.30 & 1.19$\pm$0.15 & 14$\pm$15  & 15$\pm$14   \\
\noalign{\smallskip}
LoCuSS & 4 & 5.89 & 0.24 & 0.93$\pm$0.29  & 9$\pm$20 & 22$\pm$29  \\
\noalign{\smallskip}
WtG  & 7 & 15.20 & 0.41 & 0.48$\pm$0.18  & 16$\pm$0.17  & 18$\pm$16   \\
\noalign{\smallskip}
\hline
\end{tabular}
}
\end{table}

Finally, there are 17 clusters in common with Weighing the Giants (WtG), the sample analyzed by \citet[hereafter A14]{2014MNRAS.439...48A} who reported WL masses for 51 massive clusters. We find M$_{\rm HE}$/M$_{\rm WL}$=0.61$\pm$0.12, a much larger discrepancy than the one observed with respect to the other samples. There is a clear correlation between the dynamical state (i.e., M$_{\rm par}$) and M$_{\rm HE}$/M$_{\rm WL}$ (i.e., Spearman r=0.52, p=0.03). In fact, when splitting the sample in relaxed and disturbed clusters, we find M$_{\rm HE}$/M$_{\rm WL}$=0.71$\pm$0.17 and M$_{\rm HE}$/M$_{\rm WL}$=0.48$\pm$0.18, respectively. 

The intrinsic scatter $\sigma_{\rm WL}$ of the different samples varies from $\sim$10$\%$ for CCCP/MENeaCS to $\sim$35$\%$ for APEX-SZ clusters. Also for WtG the scatter is $>$30$\%$ but is probably boosted by the clear dependence of the mass ratio on the dynamical state.  In fact, both subsamples of relaxed and disturbed clusters show a smaller scatter (i.e., 15-20$\%$), although with large uncertainties. The scatters $\sigma_{\rm WL}$ and $\sigma_{\rm HE}$ are often at the same level. An exception is CCCP/MENeaCS  which show a significantly larger $\sigma_{\rm HE}$ for the disturbed systems.

\subsection{X-ray vs caustic masses} \label{sect:cau}
An alternative technique to estimate cluster masses is through the caustic method, that makes use of galaxy dynamics. The caustic, a trumpet-shaped region in the diagram of line-of-sight velocity versus projected distance, defines which galaxies lie inside and outside a cluster assuming that a system cannot have a velocity larger than the escape velocity. Caustic-derived masses (i.e., M$_{\rm c}$) are insensitive to the physical processes that might cause hydrostatic biases and are subject to different systematic uncertainties than lensing masses.

One of the advantages of the  caustic approach is that it does not rely on the hypothesis of dynamical equilibrium. The drawback is that a reliable measurement of the underlying dark matter distribution requires a large number of member galaxies (e.g., \citealt{2013ApJ...768..116S}).

\begin{figure}[h!]
\centering
\includegraphics[width=0.45\textwidth]{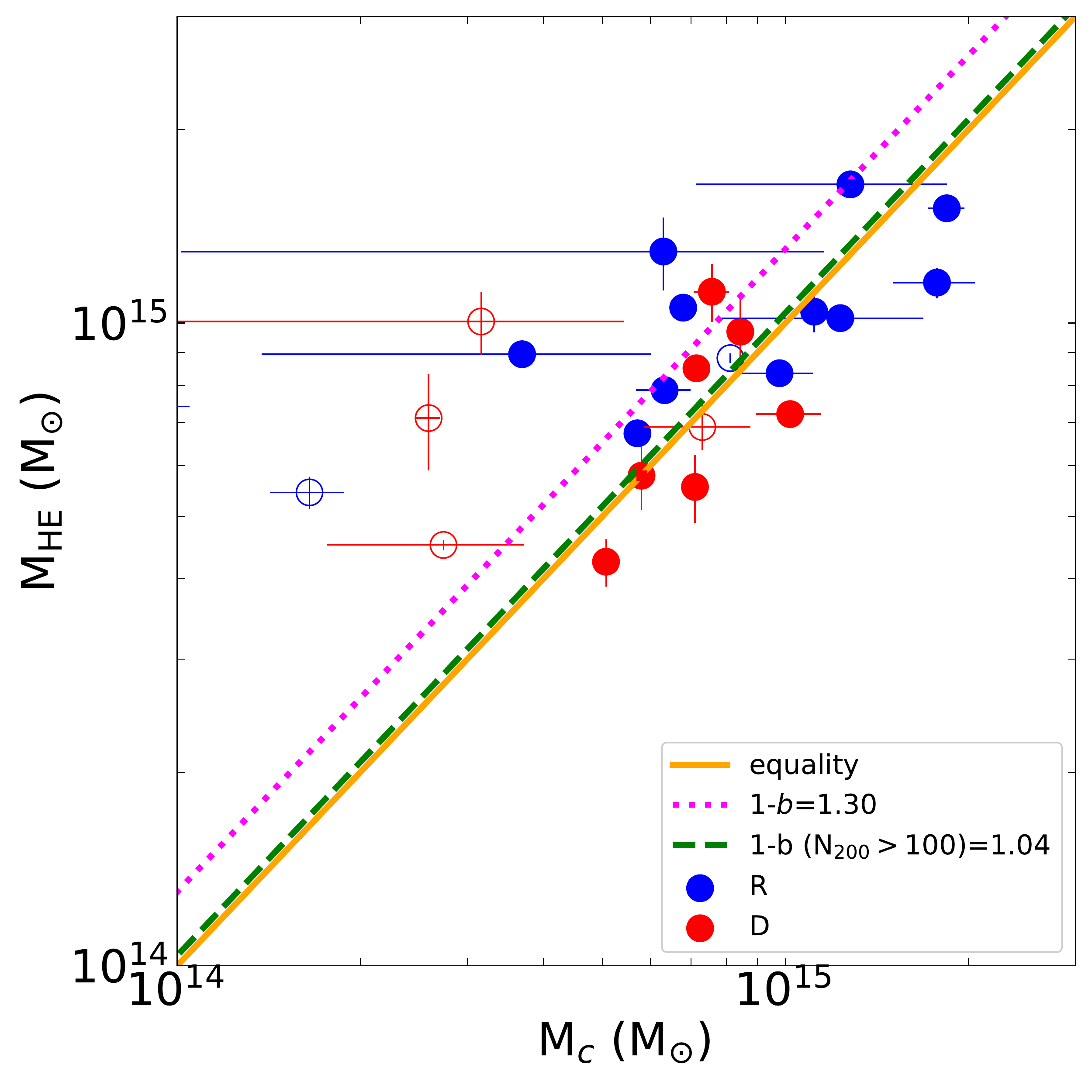}
\caption{Comparison between X-ray and caustic masses for the 25 clusters in common with \cite{2020ApJ...891..129S}. Relaxed (R) and disturbed (D) clusters are plotted with blue and red circles, respectively. Filled  and  empty  circles represent  the clusters with N$_{200}$$>$100 and N$_{200}$$<$100 , respectively. The dotted magenta line represents the average M$_{\rm HE}$/M$_{\rm c}$ ratio. The dashed green line represents the average M$_{\rm HE}$/M$_{\rm c}$ ratio obtained after removing the clusters with less of 100 galaxies within R$_{200}$.}
\label{fig:HEcau}
\end{figure}

There are 25 clusters in common with \cite{2020ApJ...891..129S} who obtained the caustic masses at R$_{200}$ for the HeCS-omnibus cluster sample (see also \citealt{2016MNRAS.461.4182M}). For a comparison with their masses, we computed the M$_{200}$ masses for the L20 sample. We found that the caustic masses are, on average, smaller than the X-ray-derived masses with no indication for a relation with the dynamical state of the clusters (see Table \ref{tab:xopt} and Fig. \ref{fig:HEcau}). However, we note that the clusters showing the largest difference between X-ray and caustic masses tend to have a small N$_{200}$, which is the number of spectroscopic members within R$_{200}$. For instance, when we split the sample based on N$_{200}$, we find M$_{\rm HE}$/M$_{\rm c}$=1.04$\pm$0.11 for clusters with N$_{200}$$>$135 and  M$_{\rm HE}$/M$_{\rm c}$=1.72$\pm$0.27 for clusters with N$_{200}$$<$135. This is also supported by the good anti-correlation between M$_{\rm HE}$/M$_{\rm c}$ and N$_{200}$ (i.e., Spearman r=-0.54, p$<$0.01). Also, the scatter significantly decreases when clusters with few member galaxies are removed, and becomes comparable with the one observed for WL samples. The low bias and relatively small scatter $\sigma_{\rm dyn}$ is found also with different cuts on N$_{200}$ and start to show significant deviations when clusters with N$_{200}$$<$100 are included in the fit.  

\begin{table}[t!]
\caption{Summary of the mass ratio between X-ray and dynamical cluster masses. The comparison is done at R$_{200}$ for the HeCS-omnibus caustic masses by \cite{2020ApJ...891..129S} and at R$_{500}$ for the HIFLUGCS masses estimate from the velocity dispersion by \cite{2017A&A...599A.138Z}. Masses are in units of 10$^{14}$M$_{\odot}$. The intrinsic scatters are given in percent.}
\label{tab:xopt}
\resizebox{0.49\textwidth}{!}{
\begin{tabular}{c c c c c c c}
\hline
\hline
\noalign{\smallskip}
\multicolumn{7}{c}{All} \\
\noalign{\smallskip}
\hline
\noalign{\smallskip}
Sample & N & $\tilde{\rm M}_{dyn}$ & $\tilde{\rm z}$ & 1-$b$ & $\sigma_{dyn}$ & $\sigma_{HE}$ \\
\noalign{\smallskip}
\hline
\noalign{\smallskip}
HeCS-omnibus & 25 & 7.10 & 0.16 & 1.30$\pm$0.14 & 76$\pm$11 & 8$\pm$8  \\
\noalign{\smallskip}
HeCS-omnibus (N$>$135) & 12 & 9.31 & 0.18 & 1.03$\pm$0.10 & 26$\pm$11 & 13$\pm$10  \\
\noalign{\smallskip}
HeCS-omnibus (N$<$135) & 13 & 5.80 & 0.14 & 1.72$\pm$0.27 & 103$\pm$20 & 18$\pm$12  \\
\noalign{\smallskip}
HeCS-omnibus (N$>$100) & 18 & 7.36 & 0.18 & 1.04$\pm$0.11 & 26$\pm$12 & 14$\pm$11  \\
\noalign{\smallskip}
HeCS-omnibus (N$<$100) & 7 & 2.74 & 0.14 & 2.34$\pm$1.48 & 121$\pm$41 & 19$\pm$15  \\
\noalign{\smallskip}
HIFLUGCS & 12 & 4.89 & 0.08 & 0.95$\pm$0.17 & 49$\pm$18 & 14$\pm$14  \\
\noalign{\smallskip}
\hline
\multicolumn{7}{c}{}\\
\hline
\hline
\noalign{\smallskip}
\multicolumn{7}{c}{Relaxed} \\
\noalign{\smallskip}
\hline
\noalign{\smallskip}
Sample & N & $\tilde{\rm M}_{dyn}$ & $\tilde{\rm z}$  & 1-$b$ & $\sigma_{dyn}$ & $\sigma_{HE}$ \\
\noalign{\smallskip}
\hline
\noalign{\smallskip}
HeCS-omnibus & 14 & 7.46 & 0.19 & 1.37$\pm$0.21 & 99$\pm$19 & 8$\pm$9  \\
\noalign{\smallskip}
HIFLUGCS & 7 & 3.77 & 0.09 & 1.28$\pm$0.23 & 37$\pm$23 & 23$\pm$15  \\
\noalign{\smallskip}
\hline
\multicolumn{7}{c}{}\\
\hline
\hline
\noalign{\smallskip}
\multicolumn{7}{c}{Disturbed} \\
\noalign{\smallskip}
\hline
\noalign{\smallskip}
Sample & N & $\tilde{\rm M}_{dyn}$ & $\tilde{\rm z}$ & 1-$b$ & $\sigma_{dyn}$ & $\sigma_{HE}$ \\
\noalign{\smallskip}
\hline
\noalign{\smallskip}
HeCS-omnibus & 11 & 7.10 & 0.11 & 1.17$\pm$0.16 & 41$\pm$15  & 25$\pm$14  \\
\noalign{\smallskip}
HIFLUGCS & 5 & 9.15 & 0.07 & 0.69$\pm$0.23 & 12$\pm$19 & 22$\pm$25  \\
\noalign{\smallskip}
\hline
\end{tabular}
}
\end{table}

\subsection{X-ray vs virial masses} \label{sect:sigma}
As an alternative to the caustic method, one can derive dynamical masses using the virial theorem (i.e., M$_{\rm v}$ hereafter). This method is based on the hypothesis of dynamical equilibrium, spherical symmetry, and on the additional assumptions that all the galaxies have the same mass and their spatial and velocity distribution follow those of dark matter particles. Despite these approximations, the virial theorem has proven to be a good estimator even for a small number of cluster galaxies (e.g., see \citealt{2006A&A...456...23B} and \citealt{2013MNRAS.430.2638M}).

\begin{figure}[h!]
\centering
\includegraphics[width=0.45\textwidth]{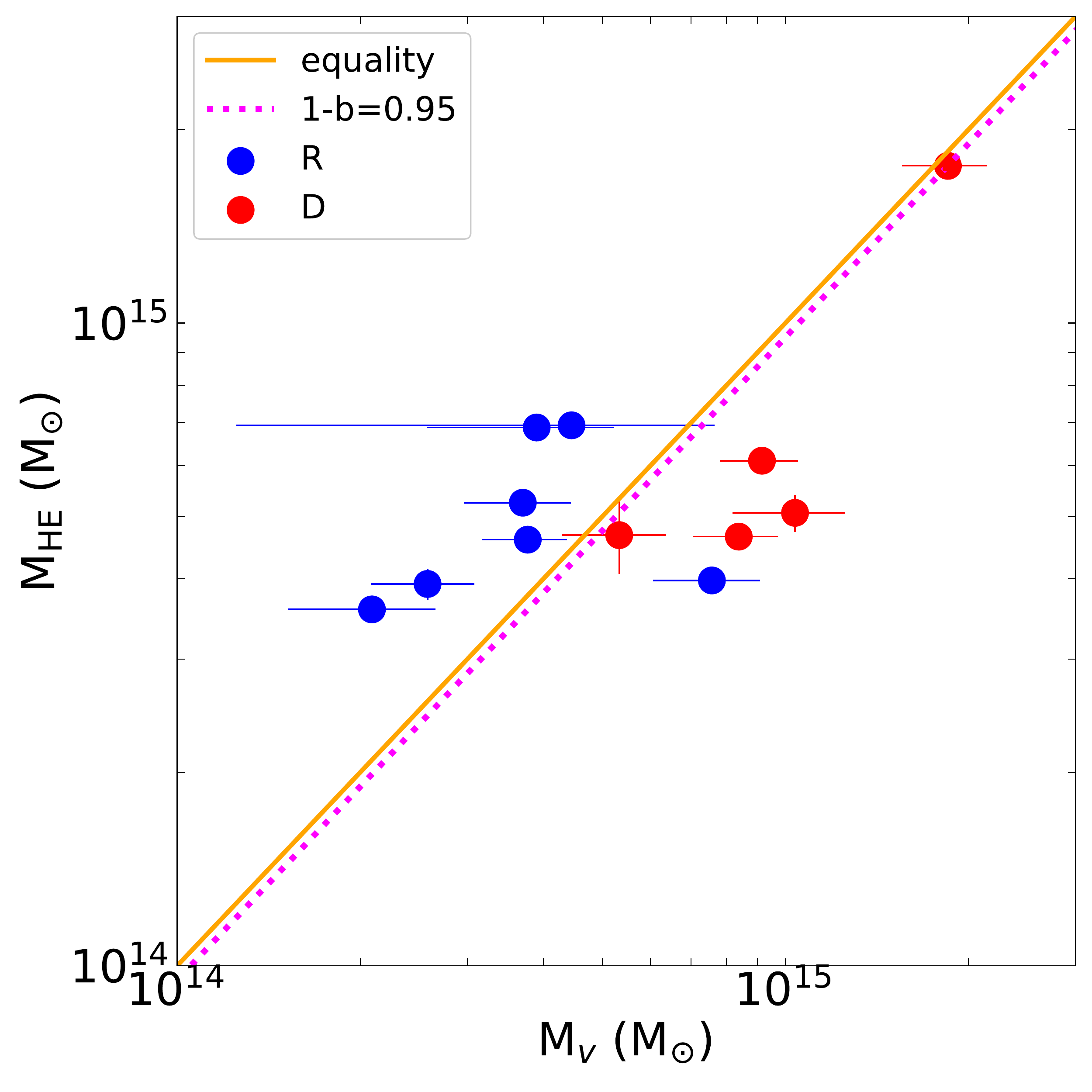}
\caption{Comparison between X-ray and virial masses for the 12 HIFLUGCS clusters in common with \cite{2017A&A...599A.138Z}. Relaxed (R) and disturbed (D) clusters are plotted with blue and red circles, respectively. The dotted magenta line represents the average M$_{\rm HE}$/M$_{\rm v}$ ratio.}
\label{fig:HEsigma}
\end{figure}

\cite{2017A&A...599A.138Z} obtained the dynamical masses for the HIFLUGCS sample based on the velocity dispersion of the member galaxies. For the 12 clusters in common with the sample in L20, we find reasonably good agreement with a mass ratio consistent with zero (i.e., 1-$b$=0.95$\pm$0.18). However, relaxed and disturbed objects clearly populate different regions of the M$_{\rm HE}$-M$_{\rm v}$ plane (see Fig. \ref{fig:HEsigma}) although larger samples are clearly needed to obtain firmer conclusions. When the HE masses are computed at R$_{500}$ provided by \cite{2017A&A...599A.138Z}, we obtain a mass ratio of 0.97$\pm$0.10 and a significant reduction of the scatter (i.e., $\sigma_{\rm dyn}$=18$\pm$12$\%$ and  $\sigma_{\rm HE}$=8$\pm$7$\%$).

\section{Discussion} \label{sect:Discussion}
\subsection{X-ray hydrostatic masses}
\cite{2014MNRAS.438...49R} and \cite{2015MNRAS.450.3633S} reported that the X-ray properties estimated by competing groups can show large discrepancies. They found that total masses can differ by a factor of two even for relaxed clusters. The major sources of the discrepancies have been ascribed to differences in the data analysis, to cross-calibration problems between instruments, and to different techniques to recover the mass. 

As seen in Sect. \ref{sectMxray} the total masses obtained in L20 are systematically lower than the masses published by other authors (at least for the ones discussed in this paper). However,  while the discrepancy can be up to 30$\%$ for individual clusters, the average difference in sample pairs is only a few percent with respect to the values by \cite{2014ApJ...794..136D} and  \cite{2011A&A...536A..11P} and $\sim$10$\%$ with respect to \cite{2019A&A...621A..39E}, \cite{2014MNRAS.443.2342M}, and \cite{2017MNRAS.469.3738S}.

In  Sect. \ref{sectMxray} we showed that part of these differences is indeed associated with the mass reconstruction methodology. Traditionally, as discussed in \cite{2013SSRv..177..119E} and \cite{2019SSRv..215...25P}, two main approaches have been used to determine the mass profile. In one case, a parametric mass model (e.g., NFW) is assumed while in the second, two functional forms are fitted to the density and temperature profiles and propagated through the HE equation to derive the mass profile. In this paper, we referred to the masses estimated with the first and second methods as M$_{\rm HE}$ and M$_{\rm for}$, respectively. Our results show that the M$_{\rm for}$ are, on average, 9$\%$ smaller than the M$_{\rm HE}$ masses. One of the reasons is a poor radial sampling (i.e., few temperature measurements) of some clusters for which the model extrapolation can diverge (see, e.g., \citealt{2018A&A...617A..64B} for a more extended discussion). In fact, when we compare the 25 clusters with at least 10 temperature bins the average difference between M$_{\rm HE}$ and M$_{\rm for}$ masses is less than 1$\%$.  When the X-ray data allow only a few temperature measurements, the assumption of the NFW profile, which is a physically motivated model and is found to provide a good fit for massive clusters, can be used to obtain mass estimates without introducing strong systematics if the temperature profile is measured at least out to 0.5R$_{500}$, as shown by \cite{2017MNRAS.471.1370S}. Our results are also in agreement with \cite{2019A&A...621A..39E} and \cite{2018A&A...617A..64B} who analyzed samples of 13 and 5 clusters with high X-ray quality data, and found that different mass estimation methods yield reasonably consistent results. 

Another cause for the lower masses in L20 is the use of the total (i.e., neutral and molecular) column density which has a significant impact on the measured temperatures in the regions with high molecular contributions. The inclusion of the molecular component allows a better fit to the X-ray data (see \citealt{2019MNRAS.483..540L} for more details) and for very high total column densities (i.e., $N_H$$>$10$^{21}$$ {\rm cm^{-2}}$) can lead to estimated temperatures (and therefore total masses) that are 10-15$\%$ lower than using only the neutral component. However, for most clusters the effect is limited to a few percent.

\begin{table}[t!]
\centering
\caption{Summary of the M$_{\rm HE}$/M$_{\rm WL}$ ratio estimated at R=1Mpc.  Masses are in units of 10$^{14}$M$_{\odot}$. The intrinsic scatters are given in percent.}
\label{tab:xwl1mpc}
\begin{tabular}{@{} l @{\hspace*{3mm}} c @{\hspace*{3mm}}c  @{\hspace*{3mm}}c @{\hspace*{3mm}}c @{\hspace*{3.5mm}}c @{\hspace*{3.5mm}}c @{\hspace*{3mm}}c @{}}
\hline
\hline
\noalign{\smallskip}
\multicolumn{7}{c}{All} \\
\noalign{\smallskip}
\hline
\noalign{\smallskip}
Sample & & N & $\tilde{\rm M}_{WL}$ & $\tilde{\rm z}$ & 1-$b$ & $\sigma_{WL}$ & $\sigma_{HE}$ \\
\noalign{\smallskip}
\hline
\noalign{\smallskip}
LC$^2$-single & All & 62 & 8.37 & 0.21 & 0.83$\pm$0.04 & 21$\pm$4 & 22$\pm$5  \\
\noalign{\smallskip}
LC$^2$-single & R & 34 & 8.65 & 0.21 & 0.84$\pm$0.05 & 20$\pm$6 & 14$\pm$6  \\
\noalign{\smallskip}
LC$^2$-single & D & 28 & 8.09 & 0.29 & 0.83$\pm$0.08 & 23$\pm$10 & 30$\pm$7  \\
\noalign{\smallskip}
\hline
\end{tabular}
\end{table}

\subsection{{\rm M$_{\rm HE}$-M$_{\rm SZ}$ and M$_{\rm SZ}$-M$_{\rm WL}$}}
The PSZ2 masses were estimated from the M-Y$_{\rm SZ}$ calibrated using the M-Y$_{\rm X}$ relation by \cite{2010A&A...517A..92A} which links the hydrostatic mass to the Y$_{\rm X}$ parameter, the X-ray equivalent of the integrated Compton parameter Y$_{\rm SZ}$. Considering uncertainties and scatter, the M-Y$_{\rm X}$ relation for the ESZ sample obtained in L20 is in good agreement with the one by \cite{2010A&A...517A..92A} in particular in the low mass regime. However, in the high mass regime, the difference is about 5-6$\%$. Therefore, the difference of 4$\%$ that we observe at the same radius (i.e., R$_{500}$ from {\it Planck}) can be easily explained by the small offset in the M-Y$_{\rm X}$ relation. 

We found a mild mass-dependent trend between X-ray and {\it Planck} SZ masses as previously obtained by other studies. \cite{2017MNRAS.469.3738S} suggest that this effect is caused by the {\it Planck} selection function and the intrinsic scatter of the hydrostatic mass estimates. However, since the {\it Planck} masses are obtained using the M-Y$_{\rm X}$ relation, it is also possible that there is a mass dependent effect introduced by the calibration of the scaling relations.

The agreement between the X-ray and SZ masses is better for relaxed clusters (4$\pm$3$\%$)  than disturbed clusters (11$\pm$3$\%$). Since the relation used to estimate the {\it Planck} masses is the same, and there is no significant difference in the M-Y$_{\rm X}$ for relaxed and disturbed systems (e.g., see L20), the larger M$_{\rm HE}$/M$_{\rm SZ}$ ratio for disturbed systems may indicate that the SZ signal is somehow impacted (i.e., overestimated). For instance, due to the large {\it Planck} beam, the integrated SZ signal may include unidentified substructures along the line of sight, which instead could be masked in the X-ray analysis. Projection effects are also a source of intrinsic scatter (e.g., \citealt{2012MNRAS.419.1766K}), and indeed we note that the scatter in the low mass regime is higher than the one for massive clusters (10$\%$ vs 4$\%$). 
 
\cite{2017MNRAS.472.1946S}, for a subsample of 35 PSZ2 clusters, found that the {\it Planck} masses are biased low with respect to WL masses by 27$\pm$11(stat)$\pm$8(sys)$\%$. Instead, the bias obtained with the heterogeneous LC$^2$ catalog is $\sim$13$\%$ (see  \citealt{2015MNRAS.450.3649S}). For the 62 ESZ clusters in common with LC$^2$, we estimate a mass ratio of 1-$b$=0.80$\pm$0.06 (corresponding to a bias of $b$=20$\pm$6$\%$), essentially in agreement with previous findings.  Due to the large uncertainties, it is hard to say if there is any trend with the dynamical state, although, based on the average values, disturbed clusters show a slightly better agreement between the SZ and WL masses (i.e., 1-$b$=0.78 and 1-$b$=0.83 for disturbed and relaxed clusters, respectively). This is in agreement with the above-mentioned suggestion of an overestimated SZ signal for disturbed clusters. However, another possibility is that the WL masses are on average slightly underestimated.

\subsection{{\rm M$_{\rm HE}$ and M$_{\rm WL}$ comparison}}
\begin{figure}[t!]
\includegraphics[width=0.48\textwidth]{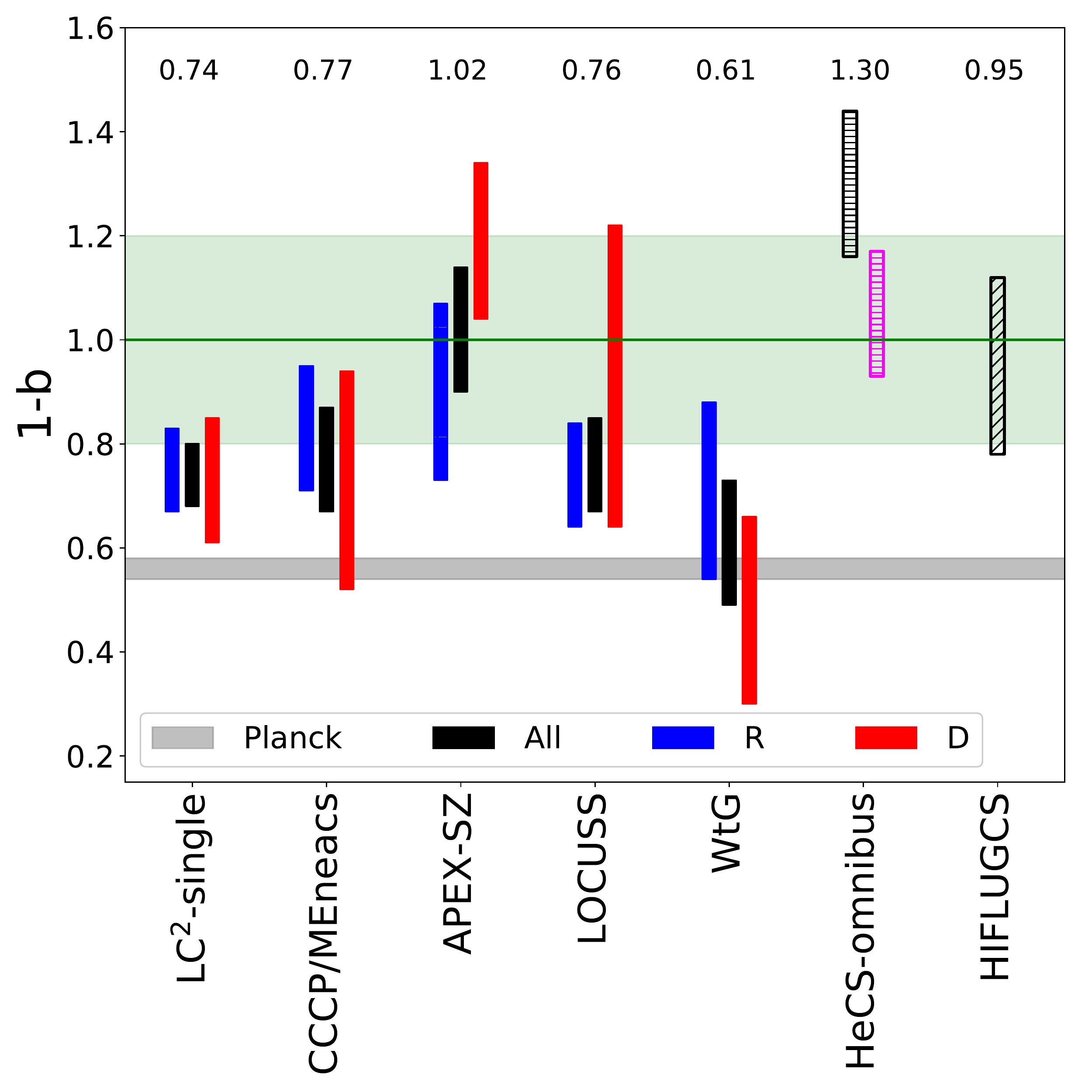}
\caption{Hydrostatic mass bias for each of the subsamples investigated in this paper. Filled color bars represent the mass bias estimated using weak-lensing masses, while horizontal and diagonal hatches show the estimates using masses from caustics and velocity dispersion, respectively. In magenta we show the bias obtained with caustics after removing the clusters with few member galaxies.
The bar lengths enclose the 1$\sigma$ estimates and the central values are also reported on the top of the panel. The grey line represents the bias necessary to reconcile {\it Planck} number counts and CMB. The green shaded region indicates the area defined as $\pm$20$\%$ centered on 1-$b$=0.}
\label{fig:summary}
\end{figure}

We compared our estimates of the hydrostatic mass with constraints obtained with weak-lensing analysis from different groups (see Sect. \ref{sect:xwl}). The mass ratio obtained with the LC$^2$ compilation is 1-$b$=0.74$\pm$0.06 (see Table \ref{tab:xszwl} and Figure \ref{fig:summary}). This is consistent with predictions from hydrodynamical simulations (e.g. \citealt{2012NJPh...14e5018R}, \citealt{2016ApJ...827..112B}).  Moreover, as discussed by \cite{2015MNRAS.450.3633S}, given the relation between mass and overdensity-radius, mass differences are inflated when computed at R$_{500}$.
Therefore, we also compare the masses obtained within a physical aperture of 1 Mpc (see Table \ref{tab:xwl1mpc}), keeping in mind that this aperture refers to a physically different region in clusters with different mass. The average difference decreases by $\sim$9$\%$ with also a substantial reduction of the intrinsic scatter. 

The large differences between the total masses in L20 and the LC$^2$ sample is partially driven by the larger mass ratio (see Table \ref{tab:xszwl} and Figure \ref{fig:summary}) observed with WtG (masses from A14) and CLASH (masses from \citealt{2016ApJ...821..116U}) programs, whose results agree. In fact, of the 62 matched clusters, 18 of the matched clusters are included either in WtG or CLASH.

The mass ratio observed, when considering the WL masses from A14, is 1-$b$=0.61$\pm$0.12, which corresponds to a bias of 39$\pm$12$\%$. This is consistent with the value required to reconcile {\it Planck} cluster number counts and {\it Planck} primary CMB mesurements. Our result is in agreement with the finding by \cite{2014MNRAS.443.1973V} who found an overall mass ratio of M$_{\rm Planck}$/M$_{\rm WtG}$=0.69$\pm$0.07.  Also,  although the uncertainties are large, this is the only sample for which the hydrostatic bias for disturbed systems is significantly higher than the bias obtained for relaxed systems, which is what would be expected if the WL masses are indeed unbiased and the hydrostatic bias is larger for dynamical active clusters.

The other samples show a higher mass ratio (corresponding to a lower hydrostatic bias,  i.e., 20-30$\%$) and a less clear dependence on the dynamical state.  The results from K19 show an anti-correlation even though not statistically significant. Since, relaxed clusters are expected to have a smaller mass bias than disturbed objects (e.g., \citealt{2016ApJ...827..112B}), the lack of correlation of the mass ratios with the dynamical state may suggest that some systematics in the WL analyses are still not accounted for. For instance,  it is worth noting that WL masses can be significantly underestimated because of massive sub-clumps (\citealt{2010A&A...514A..93M}) or uncorrelated  large-scale matter projections along the line of sight (\citealt{2011ApJ...740...25B}).
For instance, the lensing masses are typically overestimated when the cluster is elongated along the line of sight, and underestimated when it is elongated on the plane of the sky. If we focus on the more relaxed clusters, we find that the mass ratios, derived from the different subsamples, tend to agree better, and correspond to biases of the order of 5-20$\%$. Instead, when disturbed clusters are considered, the mass ratios vary a lot from sample to sample. One of the reasons is that each of the lensing subsamples contains only a small fraction of disturbed clusters (e.g., only 4 in O16). This makes possible for the lensing bias of individual systems to not be fully averaged out. We also stress that, although the subdivision between relaxed and disturbed clusters using M$_{\rm par}$, is only relative, SZ surveys are thought to be very close to being mass selected, and therefore the M$_{\rm par}$ distribution is probably indicative of the cluster population in the Universe.  

We investigated any dependence of the mass ratio on cluster mass and redshift (see Sect. \ref{sect:xwl}). We do not find any redshift dependence of the mass ratio but our clusters span a relatively small range of redshifts and we cannot place conclusive constraints. Instead, we find a moderate mass dependence, with the most massive clusters showing a larger bias in agreement with the finding by \cite{2014MNRAS.443.1973V} and \cite{2019A&A...621A..40E}, while \cite{2015MNRAS.450.3649S} and \cite{2017MNRAS.468.3322S} found the bias to be redshift rather than mass dependent. Naively, one would expect the more massive objects to be on average more disturbed than the least massive clusters, and therefore with a large bias. However, in L17, we did not find any correlation between the different morphological parameters (i.e., a measure of the dynamical state) and the total mass (see also \citealt{2010A&A...514A..32B}). This is confirmed by the Spearman rank test for the full L20 sample (r=0.02, p=0.80) and for the subsample of 62 clusters with WL masses (r=-0.09, p=0.46).

\subsection{{\rm M$_{\rm HE}$ and M$_{\rm c}$ comparison}}
In addition to the widely used WL technique to constrain the hydrostatic mass bias, we also compared the X-ray masses with caustic masses. The comparison with caustic masses determined at R$_{200}$  by \cite{2020ApJ...891..129S} required a significant extrapolation of the X-ray data, so our finding should be treated with caution. We find that the caustic masses are significantly lower than the X-ray masses in agreement with the finding by \cite{2019A&A...621A..39E}, \cite{2016MNRAS.461.4182M}, and \cite{2015MNRAS.450.3649S}.   However, \cite{2013ApJ...768..116S} found that only with a large number of spectroscopic members can the velocity field be properly sampled and return a correct mass. In particular, they showed that for N=50 the mass estimate is biased low by $\sim$20$\%$, while for N=200 the bias is consistent with zero.
After removing the clusters with fewer member galaxies, the agreement between X-ray and caustic measurements is good. Moreover, there is no indication of any dependence of the mass ratio on the dynamical state of the clusters, supporting the finding by \cite{2016MNRAS.461.4182M}. This, together with the overall trend observed with the comparison to WL masses,  suggests that there are still unknown systematics that introduce significant scatter in the relations that prevent a complete and proper estimate of the hydrostatic mass.

Previous comparisons of caustic masses with X-ray masses by \cite{2016MNRAS.461.4182M} and with {\it Planck} masses (calibrated with hydrostatic X-ray masses) by \cite{2016ApJ...819...63R} also show little or no bias. Instead, \cite{2019A&A...621A..39E} found that the caustic masses of 6 X-COP clusters, that are nearby massive objects for which the determinations of caustics might be more problematic, are significantly underestimated with respect to the hydrostatic masses. However, given the large scatter observed in the M$_{\rm HE}$-M$_{\rm c}$ relation, a comparison with a much larger sample is required to obtain more conclusive results.

\subsection{{\rm M$_{\rm HE}$ and M$_{\rm v}$ comparison}}
The velocity dispersion of member galaxies provides another way to estimate cluster masses under the hydrostatic equilibrium assumption. We find agreement when the L20 X-ray masses are compared to the masses estimated based on the velocity dispersion. In fact, for the 12 clusters in common with \cite{2017A&A...599A.138Z} we find 1-$b$=0.95$\pm$0.17 in agreement with the finding by \cite{2019A&A...621A..39E}. On the contrary, \cite{2017ApJ...844..101A} reported a bias of 1-$b$=0.64$\pm$0.11 for a sample of 17 {\it Planck} clusters. However, it is still debated whether the velocity dispersion of the galaxies is a good tracer for the velocity dispersion of the dark matter particles. In fact, differences like the so-called velocity bias, have been reported to be at the $\pm$10$\%$ level (e.g., \citealt{2013MNRAS.430.2638M}, \citealt{2018MNRAS.474.3746A}), and could partially explain the different hydrostatic bias derived in different studies. Although, there is no general agreement on the magnitude or even the sign of the velocity bias, \cite{2016ApJ...819...63R}, using a sample of 123 clusters, found that the {\it Planck} masses are not significantly biased compared to dynamical mass estimates unless, there is a significant velocity bias. We also note that our results show a significant difference between the mass ratio determined for relaxed and disturbed clusters. Although the number of objects in each subsample is small, this systematic difference suggests that the hydrostatic equilibrium assumption does affect quite differently the M$_{\rm HE}$ and M$_{\rm v}$ masses, and therefore that some systematics are not accounted for in at least one of the two methods. 

The dynamical masses have large scatter (see Table \ref{tab:xopt}) which prevents robust determination with small samples. However, the scatter for the caustic is significantly reduced when clusters with a limited number of member galaxies are excluded from the sample.

\section{Summary and Conclusion} 
We have compared the masses obtained with X-ray, WL, and dynamical analyses of a {\it Planck}-selected sample of clusters. Our main findings are the following: 

\begin{itemize}
\item[-] X-ray hydrostatic mass estimates with a forward approach (M$_{\rm for}$) and a backward approach (M$_{\rm HE}$) yield consistent results, as long as only the clusters with sufficiently large number of temperature measurements in the radial range of interest are considered. Otherwise, M$_{\rm for}$ tend to be underestimated with respect to M$_{\rm HE}$.

\item[-] the X-ray masses reported by competing groups and investigated in this paper show discrepancies lower than 10\% on average, although tensions up to 30\% can be still present in single objects, as consequence of systematic differences (mainly) in the spectroscopic analysis (see also \citealt{2015MNRAS.450.3633S}).

\item[-] the X-ray masses are smaller than the {\it Planck} SZ masses, which are derived iteratively from scaling relations calibrated using masses not corrected for the hydrostatic bias. However, the difference most probably arises from a small offset between the M-Y$_{\rm X}$ relation determined for the ESZ sample and the one used to calibrate the {\it Planck} M-Y$_{\rm SZ}$ relation. Interestingly, we find that the difference is higher for the disturbed systems which could be an indication that the SZ signal is overestimated for these systems.

\item[-] based on the compilation of WL masses from different works, the estimated M$_{\rm HE}/$M$_{\rm WL}$ mass ratio is 1-$b$$\sim$0.7-0.8. This  corresponds to an hydrostatic bias of 20-30$\%$, which is smaller than that required to resolve the discrepancy between {\it Planck} cluster number counts and primary CMB. However, the bias derived with different subsamples gives significantly different results, e.g., the comparison with 17 WtG clusters yields a mass bias of $\sim$40$\%$ (which would substantially reduce the tension) while, e.g., the comparison with 19 APEX-SZ clusters points to a zero bias. Therefore, because of the intrinsic uncertainties in the WL calibrations we cannot provide a final conclusion.

\item[-] unlike the WL masses, the dynamical masses, either from caustic or velocity dispersion, favour a scenario where X-ray hydrostatic masses have little or no bias requiring a different explanation to solve the tension between SZ cluster counts and primary CMB.  However, the dynamical masses show a large scatter and are affected by some systematics (e.g., the velocity bias) which are source of uncertainty in our final result. 

\item[-] we do not observe a significant dependence of the mass ratio 1-$b$ with the dynamical state of the clusters. The only subsamples showing a clear trend are WtG (WL masses) and HIFLUGCS (dynamical masses), while for the others the effect is small or shows an opposite behaviour. This suggests that there are systematics not accounted for in the different analyses.
However, since most of the subsamples of dynamically disturbed objects have less than 10 clusters, the dependence on the dynamical state needs to be confirmed with larger samples.

\end{itemize}
 
\begin{acknowledgements}
The authors thank the anonymous referee for useful comments and suggestions that helped improve and clarify the presentation of this work. 
L.L, S.E, and M.S. acknowledge financial contribution from the contracts ASI-INAF Athena 2015-046-R.0, ASI-INAF Athena 2019-27-HH.0, ``Attivit\`a di Studio per la comunit\`a scientifica di Astrofisica delle Alte Energie e Fisica Astroparticellare" (Accordo Attuativo ASI-INAF n. 2017-14-H.0), and from INAF ``Call per interventi aggiuntivi a sostegno della ricerca di main stream di INAF". G.S.  acknowledges  support  from {\it Chandra} grant  GO5-16126X. W.R.F. and C.J. acknowledge support from the Smithsonian Institution and the {\it Chandra} High Resolution Camera Project through NASA contract NAS8-03060. F.A-S.  acknowledges  support  from {\it Chandra} grant  GO3-14131X.
\end{acknowledgements}

\bibliographystyle{aa} 
\bibliography{eszbias} 

\begin{appendix} 
\section{Mass ratio with different estimators}\label{estimators}
In addition to the estimates from LIRA, we also quantify the mass ratios by computing the geometric mean\footnote{Opposite to the arithmetic means, the geometric mean is symmetric with respect to an exchange of the numerator and denominator (i.e., $\langle$X/Y$\rangle$=$\langle$Y/X$\rangle^{-1}$).} (hereafter gmean) which provides, similarly to the median (also quoted in the table for an easier comparison with literature), an estimate of the typical mass ratio in the sample of interest (see e.g., \citealt{2016ApJ...819...36D}, \citealt{2016MNRAS.461.3794O}, \citealt{2020ApJ...890..148U}). Errors in the estimates are computed via bootstrap resampling. The results are shown in Table \ref{tab:estimator}.

Overall, the three different estimators show the same trend when comparing the results for different samples. However, the bias obtained with LIRA are on average larger than what provided by gmean and median. This suggests that some of the objects with the largest bias have also very small statistical uncertainties, and therefore have a larger weight in the fit with LIRA. To test that, we performed the fit with LIRA assuming that each cluster have the same average error (equal to the median error of the sample). The results are in very good agreement with the estimate from gmean which provides a good measure of the typical bias in the subsample of interest. Of course, a small statistical error is often a sign of a better data quality and should indeed be accounted for during the fit. Nonetheless, if any of the measurements is systematically biased, then small statistical uncertainties associated with it can have significant impact on the interpretation of the results. This is particularly true when using data from different studies because as argued by \cite{2014MNRAS.438...49R}, the errors quoted in different papers account for different sources of statistical and systematic uncertainties and they are unable to account for the variance seen in sample pairs.       

\begin{table*}
\centering
\caption{Summary of the bias between cluster masses estimated at R$_{500}$ for X-ray, SZ, and weak-lensing, estimated with gmean, meadian and LIRA.
The uncertainties for gmean and median have been computed from bootstrapping.  Masses are in units of 10$^{14}$M$_{\odot}$. }
\label{tab:estimator}
\resizebox{0.99\textwidth}{!}{
\begin{tabular}{| c | c c c c c c | c c c c c c |}
\hline\hline
\noalign{\smallskip}
\multicolumn{1}{|c}{} & \multicolumn{12}{c|}{All} \\
\hline
\noalign{\smallskip}
\multicolumn{1}{|c}{Sample} &  \multicolumn{6}{|c}{M$_{\rm HE}$/M$_{\rm WL}$} & \multicolumn{6}{|c|}{M$_{\rm HE}$/M$_{\rm SZ}$} \\
\noalign{\smallskip}
\hline
\noalign{\smallskip}
 & N & $\tilde{M}_{\rm WL}$ & $\tilde{\rm z}$ & gmean & median & LIRA & N & $\tilde{M}_{\rm SZ}$ & $\tilde{\rm z}$ & gmean & median & LIRA \\
\noalign{\smallskip}
LC$^2$-{\it single} & 62 & 8.37 & 0.21  & 0.82$^{+0.05}_{-0.05}$ & 0.86$^{+0.02}_{-0.05}$ & 0.75$^{+0.06}_{-0.06}$ & 60 & 7.56 & 0.22 & 0.93$^{+0.02}_{-0.02}$ & 0.94$^{+0.01}_{-0.03}$ & 0.93$^{+0.02}_{-0.02}$  \\
CCCP/MENeaCS & 27 & 7.40 & 0.18 & 0.85$^{+0.09}_{-0.07}$ & 0.82$^{+0.22}_{-0.15}$ & 0.77$^{+0.10}_{-0.10}$ & 25 & 7.65 & 0.18 & 0.93$^{+0.03}_{-0.03}$ & 0.90$^{+0.02}_{-0.02}$ & 0.94$^{+0.04}_{-0.04}$  \\
APEX-SZ & 19 & 8.51 & 0.28 & 1.08$^{+0.10}_{-0.10}$ & 1.20$^{+0.02}_{-0.22}$ & 1.02$^{+0.12}_{-0.12}$ & 19 & 8.76 & 0.28 & 0.95$^{+0.03}_{-0.03}$ & 0.95$^{+0.01}_{-0.01}$ & 0.95$^{+0.04}_{-0.04}$  \\
LoCuSS & 18 & 8.24 & 0.22 & 0.80$^{+0.06}_{-0.06}$ & 0.79$^{+0.25}_{-0.11}$ & 0.76$^{+0.09}_{-0.09}$  & 18 & 7.78 & 0.22 & 0.87$^{+0.03}_{-0.03}$ & 0.88$^{+0.01}_{-0.04}$ & 0.88$^{+0.04}_{-0.04}$   \\
WtG & 17 & 11.73 & 0.29 & 0.66$^{+0.08}_{-0.07}$ & 0.65$^{+0.14}_{-0.16}$ & 0.62$^{+0.12}_{-0.12}$ & 17 & 8.27 & 0.29 &0.91$^{+0.03}_{-0.03}$ & 0.89$^{+0.04}_{-0.01}$ & 0.92$^{+0.04}_{-0.04}$  \\
\noalign{\smallskip}
\hline
\noalign{\bigskip}
\hline\hline
\noalign{\smallskip}
\multicolumn{1}{|c}{} & \multicolumn{12}{c|}{Relaxed} \\
\hline
\noalign{\smallskip}
\multicolumn{1}{|c}{Sample} &  \multicolumn{6}{|c}{M$_{\rm HE}$/M$_{\rm WL}$} & \multicolumn{6}{|c|}{M$_{\rm HE}$/M$_{\rm SZ}$} \\
\noalign{\smallskip}
\hline
\noalign{\smallskip}
 & N & $\tilde{M}_{\rm WL}$ & $\tilde{\rm z}$ & gmean & median & LIRA & N & $\tilde{M}_{\rm SZ}$ & $\tilde{\rm z}$ & gmean & median & LIRA \\
\noalign{\smallskip}
LC$^2$-{\it single} & 34 & 8.65 & 0.21 & 0.84$^{+0.06}_{-0.06}$ & 0.87$^{+0.11}_{-0.05}$ & 0.75$^{+0.08}_{-0.08}$ & 32 & 7.45 & 0.21 & 0.96$^{+0.03}_{-0.03}$ & 0.94$^{+0.03}_{-0.03}$ & 0.96$^{+0.03}_{-0.03}$  \\
CCCP/MENeaCS & 17 & 7.10 & 0.17 & 0.92$^{+0.09}_{-0.09}$ & 1.05$^{+0.02}_{-0.22}$ & 0.83$^{+0.12}_{-0.12}$ & 15 & 7.78 & 0.18 & 0.96$^{+0.05}_{-0.04}$ & 0.90$^{+0.02}_{-0.02}$ & 0.96$^{+0.05}_{-0.05}$ \\
APEX-SZ & 11 & 8.75 & 0.27 & 0.96$^{+0.11}_{-0.11}$ & 1.03$^{+0.17}_{-0.19}$ & 0.90$^{+0.17}_{-0.17}$ & 11 & 8.13 & 0.27 & 0.93$^{+0.05}_{-0.05}$ & 0.94$^{+0.01}_{-0.05}$ & 0.93$^{+0.07}_{-0.07}$ \\
LoCuSS & 14 & 9.42 & 0.22 & 0.77$^{+0.07}_{-0.07}$ & 0.68$^{+0.22}_{-0.03}$ & 0.74$^{+0.09}_{-0.09}$  & 14 & 7.96 & 0.22 & 0.89$^{+0.04}_{-0.04}$ & 0.89$^{+0.02}_{-0.04}$ & 0.89$^{+0.05}_{-0.05}$ \\
WtG & 10 & 11.46 & 0.23 & 0.80$^{+0.12}_{-0.11}$ & 0.95$^{+0.07}_{-0.29}$ & 0.71$^{+0.17}_{-0.17}$  & 10 & 8.13 & 0.23 & 0.94$^{+0.05}_{-0.04}$ & 0.90$^{+0.05}_{-0.02}$ & 0.94$^{+0.06}_{-0.06}$ \\
\noalign{\smallskip}
\hline
\noalign{\bigskip}
\hline\hline
\noalign{\smallskip}
\multicolumn{1}{|c}{} & \multicolumn{12}{c|}{Disturbed} \\
\hline
\noalign{\smallskip}
\multicolumn{1}{|c}{Sample} &  \multicolumn{6}{|c}{M$_{\rm HE}$/M$_{\rm WL}$} & \multicolumn{6}{|c|}{M$_{\rm HE}$/M$_{\rm SZ}$} \\
\noalign{\smallskip}
\hline
\noalign{\smallskip}
 & N & $\tilde{M}_{\rm WL}$ & $\tilde{\rm z}$ & gmean & median & LIRA & N & $\tilde{M}_{\rm SZ}$ & $\tilde{\rm z}$ & gmean & median & LIRA \\
\noalign{\smallskip}
LC$^2$-{\it single} & 28 & 8.09 & 0.29 & 0.80$^{+0.09}_{-0.07}$ & 0.81$^{+0.06}_{-0.20}$ & 0.74$^{+0.11}_{-0.11}$ & 28 & 7.86 & 0.29 & 0.89$^{+0.03}_{-0.03}$ & 0.95$^{+0.02}_{-0.06}$ & 0.89$^{+0.03}_{-0.03}$ \\
CCCP/MENeaCS & 10 & 8.35 & 0.19  & 0.74$^{+0.15}_{-0.11}$ & 0.63$^{+0.10}_{-0.05}$ & 0.71$^{+0.23}_{-0.23}$ & 10 & 7.12 & 0.19 & 0.89$^{+0.05}_{-0.05}$ & 0.91$^{+0.06}_{-0.06}$ & 0.96$^{+0.05}_{-0.05}$ \\
APEX-SZ & 8 & 7.57 & 0.30 & 1.27$^{+0.17}_{-0.15}$ & 1.27$^{+0.30}_{-0.22}$ & 1.19$^{+0.15}_{-0.15}$  & 8 & 9.40 & 0.30 & 0.98$^{+0.05}_{-0.05}$ & 1.00$^{+0.05}_{-0.05}$ & 0.98$^{+0.06}_{-0.06}$ \\
LoCuSS & 4 & 5.89 & 0.24 & 0.93$^{+0.14}_{-0.13}$ & 1.05$^{+0.06}_{-0.23}$ & 0.93$^{+0.39}_{-0.39}$  & 4 & 6.89 & 0.24 &  0.82$^{+0.05}_{-0.05}$ & 0.82$^{+0.10}_{-0.09}$ & 0.82$^{+0.16}_{-0.16}$  \\
WtG & 7 & 15.20 & 0.41 &  0.51$^{+0.07}_{-0.06}$ & 0.50$^{+0.01}_{-0.01}$ & 0.48$^{+0.19}_{-0.19}$ & 7 & 8.27 & 0.41 & 0.86$^{+0.05}_{-0.05}$ & 0.88$^{+0.05}_{-0.12}$ & 0.89$^{+0.08}_{-0.08}$ \\
\noalign{\smallskip}
\hline
\end{tabular}
}
\end{table*}

\section{Comparing the X-ray masses from different studies}\label{Mxray}
\begin{figure}[htb!]
\includegraphics[width=0.45\textwidth]{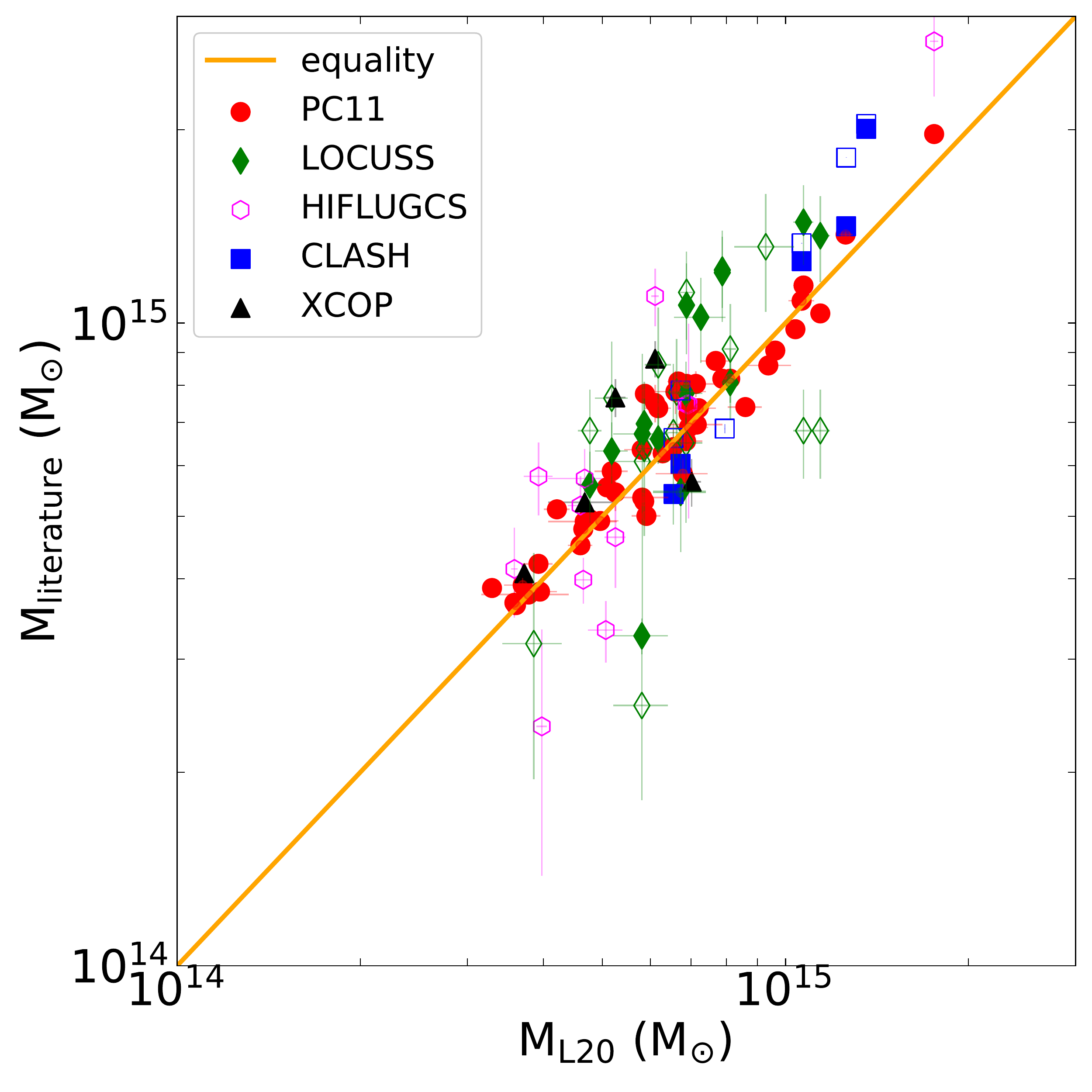}
\caption{We compare the X-ray masses obtained in L20 with the masses reported in literature. Red circles, green diamonds, magenta hexagons, blue squares, and black triangles represent the comparison with the values by \cite{2011A&A...536A..11P}, \cite{2014MNRAS.443.2342M}, \cite{2017MNRAS.469.3738S}, \cite{2014ApJ...794..136D}, and  \cite{2019A&A...621A..39E}, respectively. Fill and empty symbols represent the literature masses obtained with {\it XMM-Newton} and {\it Chandra}, respectively.}
\label{fig:Mxray}
\end{figure}

In Section \ref{sectMxray} we discussed the differences between pairs of total X-ray mass measurements performed by different groups (i.e., L20, \citealt{2011A&A...536A..11P}, \citealt{2014MNRAS.443.2342M}, \citealt{2016ApJ...819...36D}, \citealt{2017MNRAS.469.3738S}, and \citealt{2019A&A...621A..39E}). In Fig. \ref{fig:Mxray} we show the comparison  between  these  total  masses to better visualize the level of systematic differences between different studies. We note that, apart from the HIFLUGCS and a few CLASH clusters, all the plotted masses have been obtained using {\it XMM-Newton} data. Therefore, the differences cannot be attributed to cross-calibration difference between X-ray detectors, but instead to differences in the data analysis and/or to the methodology used to derive the mass (but see the discussion in Sect.  \ref{sect:Discussion} for the latter point).

\end{appendix}

%
%

\end{document}